\documentstyle[12pt]{article}

\newcommand{\ep}{\varepsilon}

\newcommand{\ra}{\rightarrow}
\newcommand{\be}{\begin{equation}}
\newcommand{\ee}{\end{equation}}
\newcommand{\prt}{\partial}
\newcommand{\bt}{\beta}
\newcommand{\lbd}{\lambda}
\newcommand{\Lbd}{\Lambda}
\newcommand{\al}{\alpha}
\newcommand{\dgr}{\dagger}
\newcommand{\dlt}{\delta}
\newcommand{\Dlt}{\Delta}
\newcommand{\vp}{\varphi}
\newcommand{\om}{\omega}
\newcommand{\Om}{\Omega}
\newcommand{\sgm}{\sigma}
\newcommand{\gm}{\gamma}
\newcommand{\Gm}{\Gamma}

\begin{document}

\begin{center}

{\Large{\bf Cooperative Electromagnetic Effects} \\ [5mm]

V.I. Yukalov, E.P. Yukalova} \\ [3mm]

{\it Joint Institute for Nuclear Research, Dubna, Russia \\
and\\
University of Western Ontario, London, Canada}

\end{center}

\vskip 2cm

\begin{abstract}

Collective phenomena in strongly nonequilibrium systems interacting with
electromagnetic field are considered. Such systems are described by
complicated nonlinear differential or integro--differential equations. The
aim of this review is to show that many nonlinear collective phenomena can
be successfully treated by a recently developed method called the Scale
Separation Approach whose name is due to the idea of separating different
characteristic space--time scales existing in nonequilibrium statistical
systems. This approach is rather general and can be applied to various
nonequilibrium physical problems, several of which are discussed here. The
problems considered not only serve as illustrations of the method but are
quite important by themselves presenting interesting physical effects, such
as Collective Liberation of Light, Turbulent Photon Filamentation,
Superradiant Spin Relaxation, Negative Electric Current, and Magnetic
Semiconfinement of Atoms.

\end{abstract}

\newpage

\section{Introduction}

Strongly nonequilibrium cooperative processes that occur in statistical
systems interacting with electromagnetic field are described by complicated
nonlinear differential and integro--differential equations. For treating
such difficult problems, a general approach has been recently developed
called the {\it Scale Separation Approach} whose basic idea is to present
the evolution equations in such a form where it could be possible to
separate several characteristic space--time scales. In many cases, different
scales appear rather naturally being directly related to the physical
properties of the considered system.

Since the scale separation approach makes the mathematical foundation for
the following applications, we start the review with presenting the basic
techniques of this approach. Then we demonstrate it by applying the method
to different physical problems related to strongly nonequilibrium processes
occurring under the interaction of electromagnetic field with matter. The
considered examples not only serve as illustrations of the method but are of
importance as such since they concern interesting and rather nontrivial
physical effects. For consideration, those effects are chosen that have been
first correctly described or predicted by the authors. Among these effects,
we would like to emphasize, as the most interesting, the following: {\it
Collective Liberation of Light, Turbulent Photon Filamentation, Superradiant
Spin Relaxation, Negative Electric Current}, and {\it Magnetic
Semiconfinement of Atoms}.

The content of the report is as follows. In Section 2 the {\it Scale
Separation Approach} is described. This method makes it possible to
solve, or to strongly simplify, many complicated systems of nonlinear
differential equations, including stochastic and partial--derivative
equations. The mathematical procedure of solving nonlinear differential
equations in the following applications is based on this approach. The
examples we consider have mainly to do with the evolution equations
describing strongly nonequilibrium statistical systems interacting with
electromagnetic fields. We concentrate our attention on collective
phenomena whose existence as such, as well as their properties, are due
to nonlinear effects. This is why we constantly have to deal with
nonlinear equations.

Resonant interactions of electromagnetic field with radiating systems are
usually describing by the Maxwell--Bloch equations, in which one often
passes to the momentum representation by means of Fourier transform.
But we prefer to work in the {\it Real--Space representation}, outlined
in Section 3, which seems to be more convenient for employing the Scale
Separation Approach. Another convenient trick we employ is the
elimination of electromagnetic field from evolution equations. For this
purpose, the operator Maxwell equations, supplemented by the Coulomb
calibration, can be rewritten in the integral form connecting the vector
potential with the retarded current formed by the radiating system.
Substituting this vector potential into evolution equations eliminates
from them electromagnetic field. In this way, we come to the system of
equations not containing explicitly electromagnetic field, instead of
which there appears an effective dipole interaction of radiating atoms.
After eliminating electromagnetic field, we have less equations, although
the price for this is that these equations become integro--differential.
nevertheless, the obtained equations are more convenient for applying to
them our method of solution. Another important advantage of the derived
equations is the possibility of taking into account quantum effects. Such
effects are often principal, while the standard semiclassical
Maxwell--Bloch equations cannot take account of them. To simplify
evolution equations, not loosing quantum effects, is the idea of the {\it
Stochastic Mean--Field Approximation} of Section 4. Since cooperative
electromagnetic phenomena are directly related to arising coherence,
Section 5 gives the definitions for {\it Dynamical Characteristics of
Coherence}.

The equations derived in the previous sections and the method of solution
developed above are applied to several concrete systems exhibiting
interesting physical properties. In Section 6, we suggest the theory of
{\t Collective Liberation of Light}, which can occur in materials with
polariton band gap. In Section 7, we consider the influence of external
fields on radiation properties of resonant atoms, checking whether it is
feasible to get {\it Amplification by Nonresonant Fields}. Section 8
discusses the so--called {\it M\"ossbauer Magnetic Anomaly} observed in
some magnetic materials. In Section 9 the {\it Problem of Pattern Selection}
is analyzed. This problem arises, for instance, when one needs to describe
resonant media with spatially nonuniform electromagnetic structures. For
treating the problem, we have suggested an original approach based on
probabilistic analysis of possible spatio--temporal patterns. This method
is applied, in Section 10, to describing {\it Turbulent Photon Filamentation}
in resonant media.

Scale Separation Approach, being a general method, can be employed for
treating strongly nonequilibrium systems of different physical nature. In
Section 11, it is used for giving a thorough picture of {\it Superradiant
Spin Relaxation} occurring in nonequilibrium nuclear magnets. This method
also makes it possible to analyse nonlinear differential equations in
partial derivatives. Such an analysis helps in finding conditions under
which unusual nonlinear effects can happen. This is illustrated in
Section 12 by describing a transient effect of {\it Negative Electric
Current} in nonuniform semiconductors. Another novel effect of {\it
Magnetic Semiconfinement of Atoms} is described in Section 13. Both these
effects have been predicted by the authors. In Section 14, we discuss
conditions when {\it Nuclear Matter Lasing} could be possible.

Throughout the review, we consider several physical systems of rather
different nature, Because of this, it is more appropriate to give all
details and to discuss the related literature in the corresponding
sections, limiting the Introduction by a brief enumeration of the
considered problems. Section 15 contains {\it Conclusion} summarizing
main results.

\section{Scale Separation Approach}

Because of the pivotal role of this approach for treating physical problems
in the following sections, we need to start by presenting its general
scheme. It is possible to separate five main steps, or parts, of the
approach: (i) stochastic quantization of short--range correlations; (ii)
separation of variables onto fast and slow; (iii) averaging method for
multifrequency systems; (iv) generalized expansion about guiding centers;
and (v) selection of scales for space structures. Below, these steps are
explicitly explained.

\vskip 5mm

{\bf 2.1. Short--Range Stochastic Quantization}. When considering
nonequilibrium processes in statistical systems, one needs to write
evolution equations for some averages $<A_i>$ of operators $A_i(t)$, where
$t$ is time and $i=1,2,\ldots,N$ enumerates particles composing the
considered system. For simplicity, a discrete index $i$ is employed,
although everywhere in what follows one could mean an operator
$A(\vec r_i,t)$ depending on a continuous space variable $\vec r_i$.

There exists the well--known problem in statistical mechanics consisting in
the fact that writing an evolution equation for $<A_i>$ one does not get a
closed set of equations but a hierarchical chain of equations connecting
correlation functions of higher orders. Thus, an equation for $<A_i>$
involves the terms as $\sum_j<A_iB_j>$ with double correlators $<A_iB_j>$,
and the evolution equations for the latter acquire the terms with triple
correlators, and so on. The simplest way for making the system of
equations closed is by resorting to the mean--field type decoupling
$<A_iB_j>\;\rightarrow\; <A_i><B_j>$. When considering radiation processes,
this decoupling is called the semiclassical approximation. Then the term
$\sum_j<A_iB_j>$ reduces to $<A_i>\sum_j<B_j>$, so that one can say that
$<A_i>$ is subject to the action of the mean field $\sum_j<B_j>$. The
semiclassical approximation describes well coherent processes, when
long--range correlations between particles govern the evolution of the
system, while short--range correlations, due to quantum fluctuations, are
not important. However, the latter may become of great importance if there
are periods of time when the long--range correlations are absent. For
example, this may happen at the beginning of a nonequilibrium process when
long--range correlations have had yet no time to develop. Then neglecting
short--range correlations can lead to principally wrong results for the
whole dynamics.

To include the influence of short--range correlations, the semiclassical
approximation can be modified as follows:
\begin{equation}
\label{1}
\sum_j <A_iB_j > = < A_i>\left ( \sum_j <B_j> + \xi \right ) \; ,
\end{equation}
where $\xi$ is a random variable describing local short--range correlations.
It is natural to treat $\xi$ as a Gaussian stochastic variable defined by
its first, $\ll\xi\gg$, and second, $\ll |\xi|^2\gg$, moments. According to
the short--range character of local fields, we should set
\begin{equation}
\label{2}
\ll \xi \gg \; = \; 0 \; .
\end{equation}
The second moment, aiming at taking into account incoherent local
fluctuations, can be defined by means of the following reasoning. Consider
the equality
$$
\ll \left | \sum_j <A_iB_j>\right |^2 \gg \; = |<A_i>|^2 \left ( \left |
\sum_j <B_j>\right |^2 + \; \ll|\xi|^2\gg\; \right )
$$
resulting from definitions (1) and (2). On the other hand, wishing to take
into account both long--range coherent as well as short--range incoherent
terms, one should write
$$
\left | \sum_j <A_iB_j>\right |^2 = |<A_i>|^2 \left ( \left | \sum_j
<B_j>\right |^2 + \sum_j |<B_j>|^2 \right ) \; ,
$$
where the first term in the brackets corresponds to the coherent while the
second term, to incoherent parts. Comparing the latter two equalities, we
come to the conclusion that
\begin{equation}
\label{3}
\ll |\xi|^2\gg \; = \sum_j |< B_j >|^2 \; .
\end{equation}
As far as short--range correlations and fluctuations are often due to
quantum effects, the manner of taking them into account by introducing a
stochastic variable $\xi$ can be named the stochastic quantization. Then the
decoupling (1) may be termed the {\it stochastic mean--field approximation}. A
similar kind of approximation has been used for taking account of
quantum spontaneous emission of atoms in the problem of atomic superradiance
[1]. Somewhat related ideas have also been used in the stochastic
quantization of quantum field theory [2].

\vskip 5mm

{\bf 2.2. Classification of Function Variations}. Employing the stochastic
mean--field approximation makes it possible to write down a closed set of
stochastic differential equations. The next step is to find such a change
of variables which results in the possibility of separating the functional
variables onto fast and slow. Let us consider, first, the variation of
functions in time. Assume that we come to the set of equations of the form
\begin{equation}
\label{4}
\frac{du}{dt} = f\; , \qquad \frac{ds}{dt} = \ep g\; ,
\end{equation}
in which $f=f(\ep,u,s,\xi,t),\; g=g(\ep,u,s,\xi,t)$, and $\ep\ll 1$ is a
small parameter. Equations (4) are complimented by initial conditions
\begin{equation}
\label{5}
u=u_0\; , \qquad s=s_0 \qquad (t=0) \; .
\end{equation}
Here, for simplicity, we deal with only two functions, $u$ and $s$, and one
small parameter $\ep$. The whole procedure is straightforwardly applicable
to the case of many functions and several parameters.

Let the functions $f$ and $g$ be such that
\begin{equation}
\label{6}
\lim_{\ep\ra 0} f\neq 0\; , \qquad \lim_{\ep\ra 0} \ep g = 0 \; .
\end{equation}
Then from Eqs. (4) it follows that
\be
\label{7}
\lim_{\ep\ra 0}\; \frac{du}{dt}\neq 0\; , \qquad
\lim_{\ep\ra 0}\; \frac{ds}{dt} = 0 \; .
\ee
This permits us to classify the solution $u$ as fast, compared to the slow
solution $s$. In turn, the slow solution $s$ is a {\it quasi--invariant}
with respect to the fast solution $u$. Thus, we may classify the functions
representing the sought solutions onto fastly and slowly varying in time.

In the case of partial differential equations, one has, in addition to time,
a space variable $\vec r$. Then the notion of fast and slow functions can be
generalized as follows [3]. Let $\vec r\in{\bf V}$, with $V$ being the
measure of the volume ${\bf V}$, and let $t\in[0,T]$, where $T$ can be
infinite. If one has
\be
\label{8}
\lim_{\ep\ra 0}\; \ll \frac{1}{V}\;\int_{\bf V}\; \frac{\prt u}{\prt t}\;
d\vec r \gg \; \neq 0 \; , \qquad
\lim_{\ep\ra 0}\; \ll\frac{1}{T}\;\int_0^T\;\vec\nabla u\; dt\gg\; \neq 0\; ,
\ee
while
\be
\label{9}
\lim_{\ep\ra 0}\; \ll \frac{1}{V}\;\int_{\bf V}\; \frac{\prt s}{\prt t}\;
d\vec r \gg \; = 0 \; , \qquad
\lim_{\ep\ra 0}\; \ll\frac{1}{T}\;\int_0^T\;\vec\nabla s\; dt\gg\; = 0\; ,
\ee
then the solution $u$ can be called {\it fast on average} with respect to
both space and time, as compared to $s$ that is {\it slow on average}. In
such a case, $s$ is again a quasi--invariant with respect to $u$. In general,
it may, of course, happen that one of the solutions is fast in time but slow
in space, or vice versa, as compared to another solution. Note that in the
Hamiltonian mechanics quasi--invariants with respect to time are called
adiabatic invariants [4]. A generalization of this notion to the case of both
space and time variables [3] is given by definition (9).

\vskip 5mm

{\bf 2.3. Multifrequency Averaging Technique.} Let us continue considering
the ordinary differential equations (4). The generalization to the case of
partial differential equations can be done similarly to the way discussed
at the end of the previous section. After classifying the function $u$ as
fast and $s$ as slow, we may resort to the Krylov--Bogolubov averaging
technique [5] extended to multifrequency systems.

Since the slow solution $s$ is a quasi--invariant for the fast variable $u$,
one considers the equation for the fast function, with the slow one kept
fixed,
\be
\label{10}
\frac{\prt X}{\prt t} = f(\ep,X,z,\xi,t) \; ,
\ee
here $s=z$ being treated as a constant parameter. The initial conditions for
Eq. (10) is
\be
\label{11}
X=u_0 \qquad (t=0) \; .
\ee
The pair of solutions
\be
\label{12}
X=X(\ep,z,\xi,t) \; , \qquad z=const
\ee
are called the {\it generating solutions}. Substituting the solution $X$
into the right--hand side of the equation for the slow function $s$, one
defines the average
\be
\label{13}
\overline g(\ep,z) \equiv \; \ll \frac{1}{\tau} \; \int_0^\tau \;
g(\ep, X(\ep,z,\xi,t), z,\xi,t) \; dt \gg \; ,
\ee
in which $\tau$ is the characteristic time of fast oscillations. In many
cases, it is sufficient to set $\tau\ra\infty$. In this way, we come to the
equation
\be
\label{14}
\frac{dz}{dt} = \ep\; \overline g(\ep,z) \; ,
\ee
with the initial condition
\be
\label{15}
z=s_0 \qquad (t=0) \; .
\ee
The solution to Eq. (14),
\be
\label{16}
z=z(\ep,t) \; ,
\ee
is to be substituted into $X$ yielding
\be
\label{17}
y(\ep,\xi,t) = X(\ep,z(\ep,t),\xi,t) \; .
\ee
Generating solutions (12) are the first crude approximations one starts with.
More elaborate solutions (16) and (17) are termed {\it guiding centers}.

Notice two points that difference the considered way of obtaining the guiding
centers (16) and (17) from the standard averaging method [5]. The first point
is in retaining in Eq. (10) the small parameter $\ep$, which makes it
possible to correctly take into account important physical effects, such as
attenuation. The  standard manner of defining the generating solutions with
setting $\ep=0$ would result in essentially more rough approximations. The
second difference is in the occurrence of the stochastic average in definition
(13), since here we are dealing with stochastic differential equations.

\vskip 5mm

{\bf 2.4. Generalized Asymptotic Expansion}. The generating solutions (12)
play the role of the trial zero--order approximation, while the guiding
centers (16) and (17) essentially improve the trial approximations.
Higher--order corrections may be obtained by presenting the general solutions
as asymptotic expansions about the guiding centers. Then, $k$--order
approximations are written as
\be
\label{18}
u_k = y(\ep,\xi,t) + \sum_{n=1}^k\; y_n(\ep,\xi,t)\; \ep^n \; , \qquad
s_k = z(\ep,t) + \sum_{n=1}^k \; z_n(\ep,\xi,t)\;\ep^n \; .
\ee
Such series are named {\it generalized asymptotic expansions} [6], since the
expansion coefficients depend themselves on parameter $\ep$. The right--hand
sides of Eqs. (4) are also to be expanded about the guiding centers yielding
$$
f(\ep,u_k,s_k,\xi,t) \simeq f(\ep,y,z,\xi,t) +
\sum_{n=1}^k\; f_n(\ep,\xi,t)\; \ep^n \; ,
$$
\be
\label{19}
g(\ep,u_k,s_k,\xi,t) \simeq g(\ep,y,z,\xi,t) +
\sum_{n=1}^k\; g_n(\ep,\xi,t)\; \ep^n \; .
\ee
Then, expansions (18) and (19) are to be substituted in Eqs. (4) with
equating the like terms with respect to the explicit powers of $\ep$. Thus,
in the first order, this gives
\be
\label{20}
\frac{dy_1}{dt} = f_1(\ep,\xi,t) - \overline g(\ep,z)\; X_1(\ep,\xi,t) \; ,
\qquad
\frac{dz_1}{dt} = g(\ep,y,z,\xi,t) - \overline g(\ep,z) \; ,
\ee
where
$$
X_1(\ep,\xi,t) \equiv \frac{\prt}{\prt z}\; X(\ep,z,\xi,t) \; , \qquad
z=z(\ep,t) \; .
$$
For the approximations of order $n\geq 2$, we get
\be
\label{21}
\frac{dy_n}{dt} = f_n(\ep,\xi,t) \; , \qquad
\frac{dz_n}{dt} = g_n(\ep,\xi,t) \; .
\ee
The initial conditions for all $n=1,2,\ldots$ are
\be
\label{22}
y_n = z_n = 0 \qquad (t=0) \; .
\ee
The functions $f_n$ and $g_n$ depend on $y_1,\; y_2,\ldots,\; y_n$, and on
$z_1,\; z_2,\ldots,\; z_n$, but it is important that the dependence on $y_n$
and $z_n$ is linear. The latter follows from the fact that expanding a
function
$$
f\left ( y + \sum_{n=1}^k\; y_n\; \ep^n\right ) = \sum_{n=1}^k f_n\;\ep^n
$$
in powers of $\ep$, one has
$$
f_1 = f'(y) y_1 \; , \quad f_2 = \frac{1}{2!}\left [ f''(y) y_1 +
f'(y)y_2 \right ] \; , \quad
f_3 = \frac{1}{3!}\left [ f'''(y) y_1 + 2f''(y) y_2 + f'(y) y_3\right ] \; ,
$$
and so on. In this way, Eqs. (20) directly define $y_1$ and $z_1$, and
Eqs. (21) are linear equations, thus, being easily integrated.

Usually, one does not need the higher--order approximations since the main
physics, in the majority of cases, is already well described by the guiding
centers (16) and (17). The latter are good approximations to the exact
solutions [7] in the time interval $0\leq t\leq T_s/\ep$, where $T_s$ is a
characteristic time of the slow--solution variation. In those cases when the
higher--order approximations are important, each $k$--order approximant can
also be improved by invoking some sort of summation [8] of asymptotic series
(18), for instance, the self--similar summation [9--12].

\vskip 5mm

{\bf 2.5. Selection of Space Structures}. The solutions to differential or
integro--differential nonlinear equations in partial derivatives are
generally nonuniform in space exhibiting the formation of different spatial
structures. And it often happens that a given set of equations possesses
several solutions corresponding to different spatial patterns or to different
scales of such patterns [13]. When there is a family of solutions describing
several possible patterns, the question arises which of these solutions, and
respectively patterns, is preferable and in what sense could it be preferable.
This problem of pattern selection is a general and very important problem
constantly arising in considering spatial structures. In this subsection we
delineate a simple way that in many cases helps to solve the problem of
pattern selection. A more refined theory will be presented in sections 9
and 10.

Assume that the obtained solutions describe spatial structures that can be
parametrized by a multiparameter $\bt$, so that the $k$--order approximations
$u_k(\bt,\vec r,t)$ and $s_k(\bt,\vec r,t)$ include the dependence on $\bt$
whose value is, however, yet undefined. To define $\bt$, and respectively
the related pattern, one may proceed in the spirit of the self--similar
approximation theory [14--23], by treating $\bt$ as a control function.
According to the theory [14--23], control functions are to be defined from
fixed--point conditions for an approximation cascade constructed for an
observable quantity. For the latter, one may take the average energy defined
as follows. The internal energy, which is a statistical average of the
system Hamiltonian, is a functional $E[u,s]$ of the solutions. Taking the
$k$--order approximations for the latter and averaging the internal energy
over the period of fast oscillations and over stochastic variables, one gets
the average energy
\be
\label{23}
E_k(\bt) \equiv \; \ll \frac{1}{\tau} \; \int_0^\tau \;
E[u_k(\bt,\vec r,t),s_k(\bt,\vec r,t) ] \; dt \gg \; .
\ee
For the sequence of approximations, $\{ E_k(\bt)\}$, it is possible to
construct an approximation cascade whose fixed point can be given by the
condition
\be
\label{24}
\frac{\prt}{\prt\bt} \; E_k(\bt) = 0 \; ,
\ee
from which one gets the control function $\bt=\bt_k$ defining the
corresponding pattern. According to optimal control theory, control
functions are defined so that to minimize a cost functional. The latter,
in our case, is naturally represented by the average energy (23). Hence,
when the fixed--point equation (24) has several solutions, one may select
of them that one which minimizes the cost functional (23), so that
\be
\label{25}
E_k(\bt_k) = {\rm abs}\min_\bt\; E_k(\bt) \; .
\ee
Equations (24) and (25) have a simple physical interpretation as the
minimum conditions for the average energy (23). However, one should keep in
mind that there is no, in general, such a principle of minimal energy for
nonequilibrium systems [13]. Therefore the usage of the ideas from the
self--similar approximation theory [14--23] provides a justification for
employing conditions (24) and (25) for nonequilibrium processes.

The scale separation approach presented in this section makes it possible
to solve rather complicated sets of nonlinear differential equations
describing various nonequilibrium phenomena in statistical systems. More
details on this approach can be found in Refs. [24--28].

\section{Real Space Representation}

When considering the interaction of atoms with electromagnetic fields, one
usually employs the so--called mode representation, expanding field operators
over mode wave functions [29,30]. These can be either free--mode functions,
that is plane waves, or resonator--mode functions depending on the resonator
geometry. We prefer to deal with the real--space representation because of
the following reasons: First, the evolution equations in this representation
are written in a form more convenient for analysing temporal nonstationary
behaviour of solutions. Second, it is more suitable for describing nonuniform
solutions corresponding to self--organized space structures. And third, this
representation is more appropriate for using the scale separation approach.
Since the real space representation is rarely considered in literature, it is
worth recalling in brief the derivation of the main equations in this
representation [31]. To understand the basis of the main evolution equations
is very important, for these equations will be constantly used in what follows.
One more peculiarity of the consideration below, differencing it form the
standard texts, is the comparison of the formulas for the cases of
electrodipole and magnitodipole transitions.

Let us have a system of radiators that can be atoms, molecules, nuclei, etc.
Assume that the size of a radiator, $a_0$, is small as compared to the mean
distance between them, $a$, as well as to the characteristic radiation
wavelength $\lbd$,
\be
\label{26}
\frac{a_0}{a} \ll 1 \; , \qquad \frac{a_0}{\lbd} \ll 1 \; ,
\ee
while the relation between $a$ and $\lbd$ can be arbitrary. Canonical
variables related to the electromagnetic field are the electric field
$\vec E$ and the vector potential $\vec A$, whose commutation relations
are
$$
\left [ E^\al(\vec r,t), \; A^\bt(\vec r\;',t) \right ] =
4\pi i\;c\;\dlt_{\al\bt}\;\dlt(\vec r - \vec r\;') \; ,
$$
$$
\left [ E^\al(\vec r,t), \; E^\bt(\vec r\;',t) \right ] =
\left [ A^\al(\vec r,t),\;  A^\bt(\vec r\;',t) \right ] = 0 \; ,
$$
where $c$ is the light velocity and the index $\al,\bt=1,2,3$, or $x,y,z$,
enumerate the Cartesian coordinates. The magnetic field is
$$
\vec H(\vec r,t) = \vec\nabla \times\vec A(\vec r,t) \; .
$$
To uniquely define the latter, we invoke the Coulomb gauge condition
$$
\vec\nabla\cdot \vec A(\vec r,t) = 0 \; .
$$
Here and in what follows the system of units is used where $\hbar\equiv 1$.

The radiator charges are described by the annihilation, $\psi$, and creation,
$\psi^\dgr$, field operators with the commutation relations
$$
\left [ \psi(\vec r,t),\; \psi^\dgr(\vec r\;',t)\right ]_{\mp} =
\dlt(\vec r - \vec r\;')\; , \qquad
\left [ \psi(\vec r,t),\; \psi(\vec r\;',t)\right ]_{\mp} = 0 \; ,
$$
$$
\left [ \psi(\vec r,t),\; \vec E(\vec r\;',t)\right ] =
\left [ \psi(\vec r,t),\; \vec A(\vec r\;',t)\right ] = 0 \; ,
$$
in which the indices minus or plus mean the commutators or anticommutators,
respectively, depending on the Bose or Fermi statistics of the charges.

Assume that in addition to the quantum radiation fields $\vec E$ and $\vec H$
there are classical fields $\vec E_0$ and $\vec H_0$ for which we have
$$
\vec E_0(\vec r,t) = - \vec\nabla\vp_0(\vec r,t) \; , \qquad
\vec H_0(\vec r,t) =\vec \nabla\times \vec A_0(\vec r,t) \; , \qquad
\vec\nabla \cdot \vec A_0(\vec r,t) = 0 \; .
$$
These additional fields can be due to external sources or can be created by
the matter which the radiators are inserted in.

Each radiator is also subject to the action of a scalar potential
$\vp_i(\vec r)$ representing all stationary Coulomb interactions. Thus, we
may introduce the total scalar and vector potentials
\be
\label{27}
\vp_{tot}(\vec r,t) = \vp_0(\vec r,t) + \sum_{i=1}^N\; \vp_i(\vec r) \; ,
\qquad \vec A_{tot}(\vec r,t) = \vec A_0(\vec r,t) + \vec A(\vec r,t) \; ,
\ee
where $N$ is the number of radiators. Then the local energy operator is
defined
as
\be
\label{28}
\hat H(\vec r,t) = \frac{1}{2m_0} \; \left [ i\;\vec\nabla +
\frac{e}{c}\; \vec A_{tot}(\vec r,t) \right ]^2 +
e\; \vp_{tot}(\vec r,t) \; ,
\ee
where $m_0$ is mass and $e$, charge of a particle. Omitting here the
relativistic term $e^2\vec A_{tot}^2/c^2$ and using the Coulomb calibration,
we have
$$
\hat H(\vec r,t) = -\;\frac{\nabla^2}{2m_0} + \frac{ie}{m_0\;c}\;
\vec A_{tot}(\vec r,t)\cdot\vec\nabla + e\;\vp_{tot}(\vec r,t) \; .
$$

The Hamiltonian of the system of radiators interacting with electromagnetic
field and with matter is written as the sum
\be
\label{29}
\hat H = \hat H_r + \hat H_f + \hat H_{rf} + \hat H_m + \hat H_{mf} \; ,
\ee
in which the terms represent, respectively, the Hamiltonians of radiators,
field, radiator--field interaction, matter, and matter--field interaction.
The Hamiltonian of the system of radiators is
\be
\label{30}
\hat H_r(t) = \int\psi^\dgr(\vec r,t) \left [ -\;\frac{\nabla^2}{2m_0} +
e\;\sum_{i=1}^N \;\vp_i(\vec r) \right ]\; \psi(\vec r,t) \; d\vec r \; .
\ee
This includes also the direct interaction of radiators with matter by means of
the effective scalar potentials $\vp_i(\vec r)$. The field Hamiltonian writes
\be
\label{31}
\hat H_f(t) = \frac{1}{8\pi} \int\left [ \vec E^2(\vec r,t) +
\vec H^2(\vec r,t) \right ]\; d\vec r \; .
\ee
The radiator--field interaction is described by
\be
\label{32}
\hat H_{rf}(t) = \int\psi^\dgr (\vec r,t) \left [ \frac{ie}{m_0\;c}\;
\vec A_{tot}(\vec r,t)\cdot \vec\nabla + e\;\vp_0(\vec r,t)\right ] \;
\psi(\vec r,t)\; d\vec r \; .
\ee
The Hamiltonians of matter and of matter--field interaction are to be
specified according to particular cases under consideration.

The size of a radiator, according to inequalities (26), is the smallest
characteristic length. If $\vec r_i$ is the center--of--mass of a radiator,
we shall use the notation
$$
\vec E_i(t) \equiv \vec E(\vec r_i,t) \; , \qquad \vec H_i(t) =
\vec H(\vec r_i,t) \; ,
$$
$$
\vec A_i(t) \equiv \vec A(\vec r_i,t) \; , \qquad
\vec E_{0i}(t) \equiv \vec E_0(\vec r_i,t) \; , \qquad \vec H_{0i}(t) =
\vec H_0(\vec r_i,t) \; .
$$
For $\vec r$ in the vicinity of $\vec r_i$, we may write
$$
\vp_0(\vec r,t ) \simeq -\vec r \cdot \vec E_{0i}(t) \qquad (\vec r\approx
\vec r_i) \; ,
$$
$$
\vec A_0(\vec r,t ) \simeq - \;\frac{1}{2}\;\vec r
\times \vec H_{0i}(t) \; , \qquad
\vec A(\vec r,t ) \simeq \vec A_i(t) -\; \frac{1}{2}\; (\vec r -\vec r_i)
\times \vec H_i(t) \; .
$$

The energy levels of each radiator are defined by the Schr\"odinger equation
$$
\left [ -\;\frac{\nabla^2}{2m_0} + e\;\vp_i(\vec r) \right ]
\psi_n (\vec r - \vec r_i) =  E_n\;\psi_n(\vec r - \vec r_i) \; ,
$$
where it is assumed that all radiators are identical and $\vp_i(\vec r)=
\vp(\vec r -\vec r_i)$. The eigenfunctions $\psi_n(\vec r - \vec r_i)$
form a complete orthonormal set enumerated by the indices $n$ and $i$, so that
$$
\int \psi_m^*(\vec r - \vec r_i)\;\psi_n(\vec r -\vec r_j)\; d\vec r =
\dlt_{mn}\;\dlt_{ij} \; , \qquad
\sum_{in} \psi_n^*(\vec r - \vec r_i)\;\psi_n(\vec r\;'- \vec r_i) =
\dlt(\vec r - \vec r\;') \; .
$$
With these functions, we may define the density of transition current
\be
\label{33}
\vec j_{mn}(\vec r) = - \frac{ie}{2m_0} \left [ \psi_m^*(\vec r)
\vec\nabla\psi_n(\vec r) - \psi_n(\vec r)\vec\nabla\psi_m^*(\vec r)
\right ]
\ee
and the transition current
\be
\label{34}
\vec j_{mn} = \int\vec j_{mn}(\vec r)\; d\vec r\; .
\ee
We also introduce the electric transition dipole
\be
\label{35}
\vec d_{mn} = e\int\psi_m^*(\vec r)\;\vec r\; \psi_n(\vec r) \; d\vec r
\ee
and the magnetic transition dipole
\be
\label{36}
\vec\mu_{mn} = \frac{1}{2c} \int\vec r\times\vec j_{mn}(\vec r) \; d\vec r\; .
\ee
Using the equality
$$
\vec\nabla = m_0\left [ \vec r,\; -\;\frac{\nabla^2}{2m_0} +
e\;\vp_i(\vec r) \right ] \; ,
$$
one can connect the electric transition current (34) and transition dipole
(35) as
\be
\label{37}
\vec j_{mn} = i\;\om_{mn}\;\vec d_{mn} \; , \qquad
\om_{mn} \equiv E_m - E_n \; .
\ee

The field operators can be expanded over the basis of the wave functions as
$$
\psi(\vec r,t) = \sum_n\;
\sum_{i=1}^N c_{ni}(t)\;\psi_n(\vec r - \vec r_i) \; .
$$
From the commutation relations for the field operators one has
$$
\left [ c_{mi}(t), \; c_{ni}^\dgr(t) \right ]_\mp =
\dlt_{mn}\; \dlt_{ij} \; , \qquad [c_{mi}(t), \; c_{nj}(t)]_\mp = 0 \; .
$$
The fact that each radiator is certainly in one of the states labelled by the
index $n$ is expressed by the unipolarity condition
\be
\label{38}
\sum_n c_{ni}^\dgr (t)\; c_{ni}(t) =  1 \; .
\ee
The wave functions $\psi_n(\vec r - \vec r_i)$, in agreement with
inequalities (26), are localized in a small region of the size of a radiator.
Such functions are called the localized orbitals. The localization condition
can be represented by the equality
$$
\int\psi_m^*(\vec r - \vec r_i)\; f(\vec r)\; \psi_n(\vec r -\vec r_j) \;
d\vec r = 0 \qquad (i\neq j) \; ,
$$
in which $f(\vec r)$ is a finite function.

Using the notations and conditions introduced above, we transform the
radiator Hamiltonian (30) to the form
\be
\label{39}
\hat H_r(t) = \sum_n\;\sum_{i=1}^N\; E_n\; c_{ni}^\dgr(t)\; c_{ni}(t) \; .
\ee
The radiator--field Hamiltonian (32) becomes
\be
\label{40}
\hat H_{rf}(t) = - \sum_{mn}\; \sum_{i=1}^N\; c_{mi}^\dgr(t)\;
c_{ni}(t) \left [ \vec d_{mn}\cdot\vec E_{0i}(t) + \frac{1}{c}\;
\vec j_{mn}\cdot\vec A_i(t) + \vec \mu_{mn}\cdot \vec B_i(t) \right ] \; ,
\ee
where
\be
\label{41}
\vec B_i(t) = \vec H_{0i}(t) + \vec H_i(t)
\ee
is the total magnetic field.

From definitions (34) to (36), we have
$$
\vec d^*_{mn} =\vec d_{nm} \; , \qquad \vec j_{mn}^* = \vec j_{nm} \; ,
\qquad \vec{\mu^*}_{mn} = \vec\mu_{nm} \; .
$$
Because the wave functions are usually either symmetric or antisymmetric with
respect to the spatial inversion, so that
\be
\label{42}
|\psi_n(-\vec r)| = |\psi_n(\vec r)| \; ,
\ee
then we see that $\vec d_{nn} =\vec j_{nn}=0$ but, in general,
$\vec\mu_{nn}\neq 0$.

The next approximation that is usually involved is related to the situation
when only a couple of radiator levels takes part in the considered process.
This happens when the transition frequency
\be
\label{43}
\om_0\equiv \om_{21} = E_2 - E_1 > 0
\ee
for these two levels is selected by means of an external alternating
field whose frequency is close to the transition frequency (43). In this
way, considering only two levels is equivalent to the quasiresonance
approximation. Then, it is convenient to introduce the transition operators
$$
\sgm_i^-(t) = c_{1i}^\dgr(t) \; c_{2i}(t) \; , \qquad
\sgm_i^+(t) = c_{2i}^\dgr(t) \; c_{1i}(t)
$$ and the population--difference operator
$$
\sgm_i^z(t) = c_{2i}^\dgr(t) \; c_{2i}(t) - c_{1i}^\dgr(t)\; c_{1i}(t) \; ,
$$
so that
$$
2c_{1i}^\dgr(t)\; c_{1i}(t) = 1 - \sgm_i^z(t) \; , \qquad
2c_{2i}^\dgr(t)\; c_{2i}(t) = 1 +\sgm_i^z(t) \; .
$$
The commutation relations for the introduced operators are
$$
[\sgm_i^-,\; \sgm_j^+ ] = -\dlt_{ij}\; \sgm_i^z \; , \qquad
[\sgm_i^-,\; \sgm_j^-] = [\sgm_i^+,\; \sgm_j^+ ] = 0 \; , \qquad
[\sgm_i^-,\; \sgm_j^z ] = 2\;\dlt_{ij}\; \sgm_i^- \; ,
$$
$$
[\sgm_i^+,\; \sgm_j^z] = - 2\;\dlt_{ij}\; \sgm_i^+ \; , \qquad
[\sgm_i^-,\; \vec A_j ] = [\sgm_i^-, \; \vec E_j] =
[\sgm_i^z,\; \vec A_j] = [ \sgm_i^z,\; \vec E_j] = 0 \; ,
$$
where all operators are taken at coinciding times.

With the notation
\be
\label{44}
\vec d_{21} \equiv \vec d \; , \qquad \vec\mu_{21}\equiv \vec\mu \; ,
\ee
we have $\vec d_{12}=\vec d^*,\; \vec\mu_{12}=\vec{\mu^*}$, and consequently
\be
\label{45}
\vec j_{12} = - \; i\;\om_0\; \vec d^*\; , \qquad
\vec j_{21} = i\;\om_0\; \vec d\; .
\ee

Since only the difference between level energies is measurable, one can set
$E_1=0$. Then the radiator Hamiltonian (39) reduces to
\be
\label{46}
\hat H_r(t) = \frac{1}{2}\; \sum_{i=1}^N \; \om_0 \; \left [ 1 +
\sgm_i^z(t) \right ] \; .
\ee
Everywhere in what follows we assume that electromagnetic fields acting
on a radiator do not change the classification of its energy levels. In the
other case it would be impossible to talk about quasiresonance. This implies
that the interaction energies of a radiator with fields are assumed to be
much smaller than $\om_0$. Because of the latter, the term
$$
\frac{1}{2}\; \sum_{i=1}^N \; \left [ (\vec\mu_{11} + \vec\mu_{22} ) +
(\vec\mu_{22} -\vec\mu_{11})\; \sgm_i^z(t)\right ]\cdot \vec B_i(t) \; ,
$$
entering the radiator--field Hamiltonian (40), can be neglected as
compared to Eq. (46). As a result, we obtain
\be
\label{47}
\hat H_{rf}(t) = - \; \sum_{i=1}^N\;
\left [ \frac{1}{c}\; \vec j_i(t)\cdot \vec A_i(t)
+ \vec d_i(t)\cdot \vec E_{0i}(t) +
\vec\mu_i(t)\cdot\vec B_i(t)\right ]\; ,
\ee
where the notation
$$
\vec j_i(t) = i\;\om_0\;\left [\vec d\;\sgm_i^+(t) -
\vec d^*\; \sgm_i^-(t)\right ]\; , \qquad
\vec d_i(t) =\vec d\; \sgm_i^+(t) + \vec d^*\; \sgm_i^- (t) \; ,
$$
\be
\label{48}
 \vec\mu_i(t) = \vec\mu\; \sgm_i^+(t) + \vec{\mu^*}\; \sgm_i^-(t)
\ee
is used. The Hamiltonian of the matter--field interaction can be written
analogously to the first term in Eq. (47) as
\be
\label{49}
\hat H_{mf}(t) = -\; \frac{1}{c}\; \sum_{j=1}^{N_0}\; \vec J_{mj}(t) \cdot
\vec A_j(t)\; ,
\ee
where $N_0$ is the number of particles forming the matter and $\vec J_{mj}$
is a local matter current having the structure of the operator
$\vec J_{mj}=(e/m)\vec p_j$, with $\vec p_j$ being the momentum of a
$j$--particle.

The transition between the quantum states $\psi_1$ and $\psi_2$ can be
either accompanied by the change of parity or not. Then from definitions
(35) and (36) it follows that one has one of two possibilities:
$$
\vec d\neq 0 \; , \qquad \vec\mu=0 \qquad (changed\; parity)\; ;
$$
\be
\label{50}
\vec d = 0 \; , \qquad \vec\mu \neq 0 \qquad (conserved\; parity)\; .
\ee
Thus, we actually have to deal with only one of the dipole transitions,
either with electric or with magnetic. Here we consider them in parallel
in order to compare these two cases.

\section{Stochastic Mean--Field Approximation}

Now it is necessary to write down the evolution equations for the operators
entering the total Hamiltonian (29) whose terms are given by Eqs (46), (31),
(47), and (49). The Heinserberg equations yield
\be
\label{51}
\frac{1}{c}\; \frac{\prt}{\prt t}\; \vec E(\vec r,t) = \vec\nabla\times
\vec H(\vec r,t) - \frac{4\pi}{c}\; \vec J(\vec r,t) \; , \qquad
\frac{1}{c}\; \frac{\prt}{\prt t}\; \vec A(\vec r,t) =
- \; \vec E(\vec r,t) \;,
\ee
which are, actually, the operator Maxwell equations, where the operator of
current is
\be
\label{52}
\vec J(\vec r,t) = \sum_{i=1}^N\; \left [ \vec j_i(t) -
c \; \vec\mu_i(t)\times \vec\nabla\right ] \dlt(\vec r - \vec r_i) +
\sum_{j=1}^{N_0}\; \vec J_{mj}(t)\; \dlt(\vec r -\vec r_j) \; .
\ee
For the transition operators we have
\be
\label{53}
\frac{d\sgm_i^-}{dt} = -\; i\; \om_0 \; \sgm_i^-  + \left ( k_0\;
\vec d\cdot\vec A_i - i\; \vec d\cdot\vec E_{0i} -
i\; \vec\mu\cdot\vec B_i\right )\; \sgm_i^z
\ee
for the lowering operator, where $k_0\equiv\om_0/c$, and the Hermitian
conjugate equation for the rising operator $\sgm_i^+$. For the
population--difference operator we get
\be
\label{54}
\frac{d\sgm_i^z}{dt} = - 2\;k_0\;\left ( \vec d\;\sgm_i^+ +\vec d^*\;
\sgm_i^-\right) \cdot \vec A_i +
2\; i\left (\vec d\; \sgm_i^+ - \vec d^*\; \sgm_i^- \right )\cdot
\vec E_{0i} +
2\; i\; \left ( \vec\mu\; \sgm_i^+ - \vec{\mu^*}\; \sgm_i^-\right )\cdot
\vec B_i \; .
\ee
From Eqs. (51), using the Coulomb calibration, we find the wave equation
\be
\label{55}
\left ( \vec\nabla^2 -\; \frac{1}{c^2}\; \frac{\prt^2}{\prt t^2}\right )
\vec A(\vec r,t) = -\; \frac{4\pi}{c}\; \vec J(\vec r,t) \; .
\ee
The solution of the latter has the form
\be
\label{56}
\vec A(\vec r,t) = \vec A_{vac}(\vec r,t) + \frac{1}{c}\; \int
\vec J\left (\vec r\;', t-\; \frac{|\vec r-\vec r\;'|}{c}\right )\;
\frac{d\vec r\;'}{|\vec r-\vec r\;'|} \; ,
\ee
in which $\vec A_{vac}$ is the vacuum vector potential being a solution
of the uniform wave equation. With the operator of current (52), the vector
potential (56) can be written as the sum
\be
\label{57}
\vec A = \vec A_{vac} + \vec A_{rad} + \vec A_{mat}
\ee
of the vacuum potential $\vec A_{vac}$, the radiator potential
\be
\label{58}
\vec A_{rad}(\vec r_i,t) = \sum_j\; \frac{1}{c\; r_{ij}}\;
\vec j_j\left ( t -\; \frac{r_{ij}}{c}\right ) +
\sum_j \; \frac{\vec r_{ij}}{r_{ij}^3}\times
\left ( r_{ij}\; \frac{\prt}{\prt r_{ij}} - 1 \right )\; \vec\mu_j
\left ( t - \; \frac{r_{ij}}{c}\right ) \; ,
\ee
and of the matter potential
\be
\label{59}
\vec A_{mat}(\vec r_i,t) = \sum_j \; \frac{1}{c\; r_{ij}}\; \vec J_{mj}
\left ( t - \; \frac{r_{ij}}{c}\right ) \; ,
\ee
where $\vec r_{ij}\equiv \vec r_i -\vec r_j,\; r_{ij}\equiv|\vec r_{ij}|$,
and the summation $\sum_j$ does not include the term with $j=i$.

Our aim is to derive the evolution equations for the variables
\be
\label{60}
u_i(t) \equiv \; <\sgm_i^-(t)>\; , \qquad
s_i(t) \equiv \; <\sgm_i^z(t)>\; ,
\ee
in which the angle brackets mean the statistical averaging over the radiator
degrees of freedom. For the double correlators, we shall employ the
{\it mean--field type decoupling}
\be
\label{61}
<\sgm_i^\al\sgm_j^\bt>\; =\; <\sgm_i^\al><\sgm_j^\bt> \qquad (i\neq j)\; .
\ee
The quantum effects due to self--action [29] can be taken into account by
including into the evolution equations the attenuation terms defined by
\be
\label{62}
\gm \equiv \frac{4}{3}\; k_0^3\; \left ( d_0^2 + \mu_0^2\right ) \; ,
\ee
where $d_0\equiv|\vec d|$ and $\mu_0\equiv|\vec\mu|$. More generally, one
includes the phenomenological longitudinal and transverse attenuation
parameters $\gm_1$ and $\gm_2$.

To take into account the retardation, we may remember that the action of
electromagnetic fields is characterized by the energies that are much
smaller than $\om_0$. That is, in the zero order one has
$\sgm_i^-\sim\exp(-i\om_0t)$, as follows from Eq. (53). This suggests to
treat the retardation by means of the formula
\be
\label{63}
<\sgm_j^-\left ( t -\; \frac{r_{ij}}{c}\right ) > \; = \;
u_j(t)\; \exp(i\; k_0\; r_{ij}) \; ,
\ee
which can be called the {\it quasirelativistic approximation} since in the
relativistic limit $c\ra\infty$, Eq. (63) becomes an identity.

Comparing the terms of the vector potential (58), induced by either
electrodipole or magnetodipole transitions, we notice their essential
difference. Really, averaging over angles gives
\be
\label{64}
\sum_j\; f(r_{ij})\; \vec r_{ij} = 0 \; ,
\ee
unless there is a special arrangement of radiators in space. Hence, the
vector potential induced by magnetodipole transitions, in usual conditions,
is negligibly small. Then for the averaged potential (58), we have
\be
\label{65}
<\vec A_{rad}(\vec r_i,t)> \; = i\; k_0^2\;  \sum_j \; \left (
\vec d\; \vp^*_{ij} \; u^*_j - \vec d^*\; \vp_{ij} \; u_j\right ) \; ,
\ee
where
\be
\label{66}
\vp_{ij} \equiv \; \frac{\exp(i\; k_0\; r_{ij})}{k_0\; r_{ij}} \; .
\ee

The influence of vacuum fluctuations and of matter is characterized by the
term
\be
\label{67}
\xi_i(t)\equiv k_0 \; \vec d\cdot \left [ \vec A_{vac}(\vec r_i,t) +
\vec A_{mat}(\vec r_i,t) \right ] \; ,
\ee
which we consider as a stochastic variable, whose properties are to be
defined by additional conditions.

In this way, we come to the evolution equations for the transverse variable,
$$
\frac{du_i}{dt} = - (i\; \om_0 +\gm_2)\; u_i -
i\; s_i\; \left ( \vec d\cdot\vec E_{0i} + \vec\mu\cdot\vec H_{0i}\right ) +
$$
\be
\label{68}
+ i\; k_0^3\;  s_i\; \vec d\cdot \sum_j \;
\left ( \vec d\;\vp^*_{ij} \; u^*_j -\vec d^*\; \vp_{ij}\; u_j\right ) +
s_i\; \xi_i \; ,
\ee
and for the longitudinal variable,
$$
\frac{ds_i}{dt} =
2\; i\; u^*_i\left ( \vec d\cdot\vec E_{0i} +
\vec\mu\cdot\vec H_{0i}\right ) -
2\; i\; u_i\left ( \vec d^*\cdot\vec E_{0i} +
\vec{\mu^*} \cdot \vec H_{0i}\right ) -
$$
\be
\label{69}
-2\; i\; k_0^3\;  (\vec d\; u_i^* +\vec d^*\; u_i) \cdot \sum_j \;
\left ( \vec d\; \vp^*_{ij}\; u^*_j -\vec d^*\;\vp_{ij}\; u_j\right ) -
\gm_1\; (s_i-\zeta) - 2\; (u^*_i\; \xi_i + u_i\; \xi_i^*)\; ,
\ee
where $\zeta\in[-1,1]$ is a pumping parameter. An equation for $u_i^*$ can
be obtained by the complex conjugation of Eq. (68). Another useful equation
is
$$
\frac{d|u_i|^2}{dt} = -2\;\gm_2\; |u_i|^2 + s_i\; (u_i^*\; \xi_i +
u_i\; \xi^*_i) -
i\; s_i\; u_i^*\left ( \vec d\cdot\vec E_{0i} +\vec\mu\cdot \vec
H_{0i}\right )+
$$
\be
\label{70}
+i\; s_i\; u_i\left (\vec d^*\cdot\vec E_{0i} + \vec{\mu^*}\cdot\vec
H_{0i}\right ) +
i\; k_0^3\; s_i\; \left ( u_i^*\; \vec d + u_i\; \vec d^*\right ) \cdot
\sum_j \; \left ( \vec d\; \vp^*_{ij}\; u_j^* - \vec d^*\; \vp_{ij}\; u_j
\right ) \; .
\ee
Equations (68) and (70) are basic for describing nonequilibrium collective
phenomena in radiating systems. The set of assumptions employed for deriving
these equations can be briefly named the {\it stochastic mean--field
approximation} since the mean--field type decoupling (61) was used for the
radiator correlators, but quantum effects are taken into account through
the stochastic variable (67).

\section{Dynamical Characteristics of Coherence}

One of the most important results of the cooperative behaviour of radiators
is the appearance of coherent radiation. The level of coherence of
electromagnetic fields can be described by the corresponding correlation
functions [32]. Here we introduce another characteristic of coherence, which
is convenient for considering the radiation from ensembles of radiators [33].

The energy density of the radiated electromagnetic field is
\be
\label{71}
W \equiv \frac{1}{8\pi}\; \left (\vec E^2 + \vec H^2 \right ) \; ,
\ee
where $\vec E =\vec E(\vec r,t)$ and $\vec H=\vec H(\vec r,t)$.
Differentiating Eq. (71) with respect to time, using the Maxwell equations
(51), and defining the intensity of scattering
\be
\label{72}
\frac{\prt W_s}{\prt t} \equiv \frac{1}{2}\; \left (\vec J\cdot\vec E +
\vec E\cdot\vec J\right )
\ee
and the Poynting vector
\be
\label{73}
\vec S \equiv \frac{c}{8\pi}\; \left ( \vec E\times\vec H -
\vec H\times\vec E\right ) \; ,
\ee
we obtain the continuity equation
\be
\label{74}
\frac{\prt}{\prt t}\; \left ( W + W_s\right ) + {\rm div} \vec S = 0\; .
\ee
The intensity of radiation into the unit solid angle is
\be
\label{75}
I(\vec n,t) \equiv \; <:\vec n\cdot\vec S(\vec r,t):>\; r^2 \; ,
\ee
where $\vec n\equiv\vec r/r,\; r\equiv|\vec r|$, and the colons imply the
normal ordering of operators. To accomplish the latter, one separates the
Hermitian operators into their conjugate parts, which, for instance, for
the vector potential (58) reads as
\be
\label{76}
\vec A_{rad}(\vec r,t) = \vec A^+(\vec r,t) + \vec A^-(\vec r,t) \; ,
\ee
where
$$
\vec A^+(\vec r,t) = \sum_j\; \left [
\frac{i\; k_0\;\vec d}{|\vec r-\vec r_j|} +
\frac{1+i\; k_0\; |\vec r-\vec r_j|}{|\vec r-\vec r_j|^3}\; \vec\mu \times
\left (\vec r-\vec r_j\right )\right ] \;
\sgm_j^+\left ( t -\;\frac{1}{c}\; |\vec r -\vec r_j|\right ) \; .
$$
Respectively, the electromagnetic positive and negative fields related to
Eq. (76) are
$$
\vec E_{rad} \equiv -\; \frac{1}{c}\; \frac{\prt \vec A_{rad}}{\prt t} =
\vec E^+ + \vec E^- \; , \qquad
\vec H_{rad} \equiv \vec\nabla\times\vec A_{rad} = \vec H^+ + \vec H^- \; .
$$
In the time and space derivatives, we may employ, for differentiating
$\sgm_j^\pm$, the relations
$$
\left ( \frac{1}{c}\; \frac{\prt}{\prt t} +
\frac{\prt}{\prt r_{ij}}\right )\;
\sgm_j^\pm\left ( t - \frac{r_{ij}}{c}\right ) = 0 \; , \qquad
\left ( \frac{\prt}{\prt r_{ij}} \pm i\; k_0 \right )\;
\sgm_j^\pm\left ( t - \frac{r_{ij}}{c}\right ) = 0 \; .
$$

In the wave zone, where $r\gg|\vec r_i|$ and
$|\vec r - \vec r_j| \simeq r - \vec n\cdot\vec r_j,\;(r\gg |\vec r_j|)$,
we have
\be
\label{77}
\vec A^+(\vec r,t) \simeq i\; \frac{k_0}{r}\;  \left ( \vec d +
\vec\mu \times \vec n\right ) \sum_j \; \sgm_j^+ \left ( t -
\frac{r-\vec n\cdot\vec r_j}{c}\right ) \; ,
\ee
from where
\be
\label{78}
\vec E^+ = - ik_0\vec A^+ \; , \qquad \vec H^+ = \vec n\times\vec E^+ \; .
\ee
Then in the part of the Poynting vector (73), describing the radiation from
the ensemble of radiators, one has
$$
\vec S_{rad} = \frac{c}{4\pi}\; \vec E_{rad} \times \vec H_{rad} \; ,
\qquad \vec H_{rad} =\vec n\times \vec E_{rad} \; .
$$
For the corresponding part of the radiation intensity (75), we get
\be
\label{79}
I_{rad}(\vec n,t) = \frac{cr^2}{4\pi} \; <: \vec E_{rad}^2 -
\left (\vec n\cdot\vec E_{rad}\right )^2: > \; .
\ee
Averaging the latter over stochastic variables and over fast oscillations
yields
\be
\label{80}
\overline I(\vec n,t) \equiv\; \frac{\om_0}{2\pi}\; \int_0^{2\pi/\om_0}
\ll I_{rad}(\vec n,t) \gg \; dt \; ,
\ee
the slow variables in the process of integration being kept fixed. For the
radiation intensity (79), this results in
\be
\label{81}
\overline I(\vec n,t) =\om_0\;\gm\;\sum_{ij}^N\; f_{ij}(\vec n)\;
\overline{<\sgm_i^+(t)\sgm_j^-(t)>} \; ,
\ee
where
\be
\label{82}
f_{ij}(\vec n) \equiv \frac{3}{8\pi}\; \left |\vec n\times\vec e\right |^2\;
\exp\left ( i\; k_0\; \vec n\cdot\vec r_{ij}\right )
\ee
and $\vec e=\vec d/d_0$ or $\vec\mu/\mu_0$ depending on the type of
radiation.

In the radiation intensity (81), we may separate the terms with the
coinciding and with different indices, so that $\sum_{ij} =\sum_{i=j} +
\sum_{i\neq j}$. This makes it possible to separate the radiation intensity
into the incoherent and coherent parts,
\be
\label{83}
\overline I(\vec n,t) = I_{inc}(\vec n,t) + I_{coh}(\vec n,t) \; ,
\ee
so that the incoherent radiation intensity is
\be
\label{84}
I_{inc}(\vec n,t) = \frac{1}{2}\;\om_0\;\gm\;\sum_{i=1}^N\; f_{ii}(\vec n)\;
[ 1 + s_i(t)]
\ee
while the coherent radiation intensity is
\be
\label{85}
I_{coh}(\vec n,t) = \om_0\; \gm\; \sum_{i\neq j}^N \; f_{ij}(\vec n)\;
\overline{u_i^*(t)u_j(t)} \; .
\ee
Here the equality $2\sgm_i^+\sgm_i^- =1 +\sgm_i^z$ was used. The total
radiation intensity is given by the integral
\be
\label{86}
I(t) \equiv \int\overline I(\vec n,t)\; d\Omega(\vec n) =
I_{inc}(t) + I_{coh}(t)
\ee
over solid angles. Here the incoherent part is
\be
\label{87}
I_{inc}(t) =\frac{1}{2}\; \om_0\; \gm\; \sum_{i=1}^N \; [1 + s_i(t) ]\; ,
\ee
and the coherent part is
\be
\label{88}
I_{coh}(t) = \om_0\; \gm\; \sum_{i\neq j}^N \; f_{ij}\;
\overline{u_i^*(t) u_j(t)} \; ,
\ee
where
\be
\label{89}
f_{ij} \equiv \int f_{ij}(\vec n)\; d\Omega(\vec n) \; , \qquad f_{ii}=1 \; .
\ee
Finally, the level of coherence can be defined [33] by means of the
{\it coherence coefficients}
\be
\label{90}
C_{coh}(\vec n,t) \equiv \frac{I_{coh}(\vec n,t)}{I_{inc}(\vec n,t)} \; ,
\qquad C_{coh}(t) \equiv \frac{I_{coh}(t)}{I_{inc}(t)} \; .
\ee
The radiation is mainly incoherent when $C_{coh}\ll 1$ and it is almost
purely coherent if $C_{coh}\gg 1$.

\section{Collective Liberation of Light}

A system of initially inverted atoms can, due to photon exchange, become
strongly correlated, as a result emitting a coherent pulse. This effect of
self--organization, accompanied by a coherent burst, is called the Dicke
superradiance [34].  This phenomenon is well studied for atoms in vacuum
[1,29,30], including different particular cases, such as superradiance in
two--component systems [35--37], superradiance from ensembles of three--level
molecules [1,38], two--photon superradiance [39,40], and so on (see citations
in Refs. [41]). When radiating atoms or molecules are placed in a solid, they
interact with phonons [42,43], which can lead to such interesting phenomena as
the laser cooling of solids [44,45].

When an atom is placed in a periodic dielectric structure, in which, due to
periodicity, a photonic band gap develops, then spontaneous emission with a
frequency inside the band gap can be rigorously forbidden [46,47]. This kind
of matter, where photon band gap appears because of the structure periodicity
in real space, has been called photonic band--gap materials. The photon band
gap also appears in natural dense media due to photon interactions with
optical collective excitations, such as phonons, magnons, or excitons [48,49].
One calls this type of the gap the polariton band gap since photons coupled
with collective excitations of a medium are termed polaritons.

If a single resonance atom is placed in a medium with a photon band gap, and
the atomic transition frequency lies inside this gap, then the spontaneous
emission is suppressed, which is named the localization of light [46,47].
This effect is caused by the formation of a photon--atom bound state [50--52].
When a collection of identical resonance atoms is doped into a medium with a
photon band gap, so that the atomic transition frequency is inside this gap,
then the atoms, in principle, can radiate because of the formation of a
photonic impurity band within the photon band gap [50,53--55]. A model case
of a concentrated sample, whose linear size $L$ is much smaller than the
radiation wavelength $\lbd$, has been considered for studying superradiance
near a photonic band gap [56,57], when the transition frequency almost
coincides with the frequency of the upper band edge. Here, following
Ref. [58], we study the realistic case of a sample with $\lbd\ll L$.

Assume that the localization of light occurs for a single atom with an
electric dipole transition, so that its population difference is always
$s_0=s(0)$. Considering an ensemble of resonance atoms, we resort to
Eqs. (68), (69), and (70). For simplicity, we write $u_i=u$ and $s_i=s$.
Introduce the effective coupling parameters
\be
\label{91}
g\equiv \frac{3\gm}{4\gm_2}\; \sum_j\;
\frac{\sin(k_0\; r_{ij})}{k_0\; r_{ij}}\; ,
\qquad g'\equiv \frac{3\gm}{4\gm_2}\;
\sum_j \; \frac{\cos(k_0\; r_{ij})}{k_0\; r_{ij}} \; ,
\ee
where $\gm\equiv 4k_0^3\;d_0^2/3$. In the absence of resonator imposing a
selected mode,
\be
\label{92}
g \approx g'\approx \frac{3\gm}{4\gm_2}\; \rho\; \lbd^3 \; ,
\ee
where $\rho$ is the density of resonance atoms. It is convenient to introduce
the effective frequency and effective attenuation defined, respectively, as
\be
\label{93}
\Om\equiv \om_0 +\gm_2\; g'\; s \; , \qquad
\Gm\equiv \gm_2\; ( 1 - g\; s) \; .
\ee
These expressions include the influence of local fields [59] through the
coupling parameters (91). Since the latter take into account the existence
of an ensemble of atoms, we may call $\Om$ and $\Gm$ the {\it collective
frequency} and {\it collective width}, respectively.

With these notations, Eq. (68) reduces to
\be
\label{94}
\frac{du}{dt} = - \left ( i\;\Om +\Gm\right )\; u + s\; \xi +
\gm_2\; \vec{e_d}^2\; ( g + i\;g')\; s\; u^* \; ,
\ee
where $\xi=\xi_i$ and $\vec e_d \equiv \vec d/d_0$. Equation (69) becomes
$$
\frac {ds}{dt} = -4\;\gm_2\; g\; |u|^2 - \gm_1\; (s - s_0) -
2\; (u^*\; \xi + u\; \xi^* )  -
$$
\be
\label{95}
- 2\;\gm_2\; \left [ (g + i\; g' )\left ( u^*\; \vec e_d\right )^2 +
(g - i\; g')\left ( u\; \vec{e_d}^*\right )^2 \right ] \; ,
\ee
where $\zeta=s_0$ takes into account that for a single atom the localization
of light occurs. And for Eq. (70), we have
\be
\label{96}
\frac{d|u|^2}{dt} = - 2\;\Gm\; |u|^2 + s\; \left ( u^*\;\xi + \xi^*\;
u\right )
+ \gm_2\;  s\; \left [ (g + i\;g' )\left ( u^*\;\vec e_d\right )^2 +
(g - ig')\left ( u\;\vec{e_d}^*\right )^2 \right ] \; .
\ee

Let us accept the natural inequalities
\be
\label{97}
\frac{\gm_1}{\Om}\ll 1 \; , \qquad \frac{\gm_2}{\Om} \ll 1 \; , \qquad
\left | \frac{\Gm}{\Om}\right | \ll 1 \; .
\ee
And, as always, we keep in mind that the interaction term (67) is small as
compared to the frequency $\Om$, or that $\ll\xi\gg\; =0$, which tells that
this term  is small on average. Then, according to Sec. 2, we may classify
the solution $u$ as fast while $s$ and$|u|^2$ as slow. Solving Eq. (94),
with $s$ being a quasi--invariant, we get
\be
\label{98}
u(t) =\left [ u_0 + s\int_0^t\; e^{(i\;\Om +\Gm)\;t'}\; \xi(t')\;
dt'\right ]\; e^{-(i\;\Om+\Gm)\;t} \; .
\ee
Introduce the notation
\be
\label{99}
\al\equiv \lim_{\tau\ra\infty}\; \frac{{\rm Re}}{\tau\;\Gm\; s}\;
\int_0^\tau \ll \xi^*(t)u(t)\gg \; dt \; ,
\ee
where ${\rm Re}$ means the real part and which, if $\ll\xi\gg=0$, takes the
form
$$
\al=\lim_{\tau\ra\infty}\; \frac{{\rm Re}}{\tau\;\Gm}\; \int_0^\tau dt
\int_0^t \; e^{-(i\;\Om+\Gm)(t-t')}\; \ll \xi^*(t)\xi(t')\gg \; dt'\; .
$$
When $\xi(t)$ is a stochastic variable corresponding to a stationary random
process, so that
$$
\ll \xi^*(t)\;\xi(t')\gg \; =\; \ll \xi^*(t-t')\; \xi(0)\gg \; ,
$$
then the notation (99) becomes
$$
\al=\lim_{\tau\ra\infty}\; \frac{{\rm Re}}{\tau\;\Gm}\; \int_0^\tau dt
\int_0^t\; e^{-(i\;\Om+\Gm)\;t'} \; \ll \xi^*(t')\;\xi(0)\gg \; dt'\; .
$$

Defining a new function
\be
\label{100}
w \equiv |u|^2 -\al\; s^2 \; ,
\ee
and averaging the right--hand sides of Eqs. (95) and (96) over time and over
stochastic variables we get
$$
\frac{ds}{dt} = -4\;g\;\gm_2\; w - \gm_1^*\;(s-\zeta^*) \; , \qquad
\frac{d|u|^2}{dt} = -2\; \Gm \; w \; ,
$$
where
$$
\gm_1^* \equiv \gm_1 + 4\;\gm_2\;\al \; , \qquad
\zeta^* \equiv \frac{\gm_1}{\gm_1^*}\; s_0 \; .
$$

In what follows, we assume that the quantity (99), describing the intensity
of interaction between atoms and matter, is small,
\be
\label{101}
|\al| \ll 1 \; .
\ee
To understand the structure of the atom--matter coupling $\al$, we may model
the random variable $\xi$ by the interaction of an atom with an ensemble of
oscillators as
$$
\xi(t) = \sum_\om\; \gm_\om \;\left ( b_\om\; e^{-i\om t} +
b_\om^\dgr \; e^{i\om t} \right ) \; ,
$$
where $b_\om$ and $b_\om^\dgr$ are Bose operators. Then the atom--matter
coupling is
$$
\al = \sum_\om\; \gm_\om^2 \; \left [ \frac{n_\om}{(\om-\Om)^2 +\Gm^2} +
\frac{1+n_\om}{(\om+\Om)^2 +\Gm^2} \right ] \; ,
$$
with $n_\om\equiv\;\ll b_\om^\dgr b_\om\gg$. If the coupling $\al$ is
small, then $\gm_1^*\approx\gm_1, \; \zeta^*\approx s_0$, and
$d|u|^2/dt\approx dw/dt$. Therefore, we obtain the equations
\be
\label{102}
\frac{ds}{dt} = -4\; g\;\gm_2\; w - \gm_1\; (s-s_0) \; , \qquad
\frac{dw}{dt} = -2\;\gm_2\; ( 1 - g\; s)\; w \; .
\ee

For transient times, when $t\ll \gm_1^{-1}$, Eqs. (102) can be solved
explicitly, giving
\be
\label{103}
s= -\; \frac{\gm_0}{g\gm_2}\; {\rm tanh}\left (\frac{t-t_0}{\tau_0}\right )
+ \frac{1}{g} \; , \qquad
w=\frac{\gm_0^2}{4g^2\gm_2^2}\; {\rm sech}^2 \left ( \frac{t-t_0}{\tau_0}
\right ) \; ,
\ee
where the integration constants $\gm_0=\tau_0^{-1}$ and $t_0$ are defined by
the initial conditions $u(0)=u_0$ and $s(0)=s_0$. For the radiation width
$\gm_0$, we get the equation
\be
\label{104}
\gm_0^2 = \Gm_0^2 + 4g^2\gm_2^2 \left ( |u_0|^2 - \al_0 s_0^2\right ) \; ,
\ee
where
$$
\Gm_0\equiv \gm_2 ( 1 - gs_0) \; , \qquad \gm_0 \equiv \frac{1}{\tau_0} \; ,
\qquad \al_0 \equiv \al(0) \; .
$$
For the delay time, we find
\be
\label{105}
t_0 = \frac{\tau_0}{2}\ln \left | \frac{\gm_0 - \Gm_0}{\gm_0 + \Gm_0}
\right | \; .
\ee
Introducing the critical coupling
\be
\label{106}
\al_c \equiv \frac{(1-gs_0)^2}{4g^2s_0^2} + \frac{|u_0|^2}{s_0^2} \; ,
\ee
we may rewrite the radiation width as
\be
\label{107}
\gm_0 = 2 g\; |s_0|\; \gm_2\; \sqrt{\al_c -\al_0} \; .
\ee

In the case of only one atom, we have to set $g=0$. Then Eqs. (102) give
$$
s=s_0 \; , \qquad w = (|u_0|^2 -\al_0\; s_0^2 )\; e^{-2\gm_2\;t} \qquad
(g=0) \; ,
$$
which means that the light is localized. But for an ensemble of atoms the
radiation becomes possible.

To find out what happens at large times, when $t\ra\infty$, we need to analyse
the stationary solutions of Eqs. (102). There are two pairs of such solutions:
\be
\label{108}
s_1^* = s_0 \; , \qquad w_1^* = 0
\ee
and
\be
\label{109}
s_2^* = \frac{1}{g} \; , \qquad w_2^* = \frac{\gm_1(gs_0-1)}{4g^2\gm_2} \; .
\ee
The stability analysis [58] shows that the fixed point (108) is stable for
$gs_0 < 1$ and unstable for $gs_0>1$, when the point (109) becomes stable.
When $gs_0<1$, the stationary point (108) is a stable node, while that (109)
is a saddle point. In the interval $1<gs_0\leq 1 + \gm_1/8\gm_2$, the fixed
point (108) is a saddle point, and that (109) is a stable node. For
$gs_0 > 1+\gm_1/8\gm_2$, the stationary solutions (108) correspond again to
a saddle point, while the fixed point (109) becomes a stable focus. In the
latter case, the pulsing regime of radiation is realized, with the asymptotic
period between pulses
\be
\label{110}
T_p = \frac{4\pi}{|\gm_1^2 + 8(1 - gs_0)\gm_1\gm_2|^{1/2}} \; .
\ee
However, at finite times the radiation pulses are not periodic, so that
the characteristic time (110) is an approximate period only for $t\ra\infty$.

In this way, when a single atom cannot radiate because of the localization of
light, an ensemble of atoms can emit coherent radiation, provided that the
interaction between atoms is sufficiently strong, so that $gs_0>1$. This is why
such an effect can be called the collective liberation of light. However,
this liberation is not complete but only partial since $s_2^*>0$.

\section{Amplification by Nonresonant Fields}

An essential enchancement of radiation can occur due to correlations between
radiators, which results in the emission of a coherent pulse. In order that
these correlations could be sufficiently strong, it is usually required that
the radiation wavelength would be much larger than the mean distance between
radiators. If the latter is not the case, it is hardly probable that the
self--organized coherence can develop. How would it be possible to amplify
the radiation intensity for a system of radiators whose wavelenght is smaller
than or comparable with the mean distance between them? This question is of
high importance for short--wave emission such as $x$--ray and $\gm$--ray
radiation. Coherent transient effects due to phase modulation of recoilless
$\gm$ radiation have been considered both theoretically and experimentally
[60--63]. A regenerated signal of gamma echo has been observed [64], which
is similar to photon echo in optics [65]. In the present section we explore
the conditions when {\it stationary} enchancement of short--wave radiation is
feasible, being due to external nonresonant fields. Some preliminary results
on the problem have been reported [66--68], based on simplified
models. Here the problem is considered more accurately, using the main Eqs.
(68) to (70). The latter, in the case of short--wave radiation, when the
interaction of radiators can be neglected, take the form
\be
\label{111}
\frac{du_i}{dt} = - (i\om_0 +\gm_2) u_i - is_i \vec d\cdot \vec E_{0i} \; ,
\ee
\be
\label{112}
\frac{ds_i}{dt} = 2i (u_i^* \vec d - u_i\vec d^*)\cdot \vec E_{0i}
-\gm_1( s_i -\zeta)  \; ,
\ee
\be
\label{113}
\frac{d|u_i|^2}{dt} = -2\gm_2|u_i|^2 - is_i (u_i^*\vec d - u_i\vec d^*)\cdot
\vec E_{0i} \; .
\ee
The initial conditions are $u_i(0)=u_0$ and $s_i(0)=s_0$.

Assuming, as usual, the existence of small parameters
\be
\label{114}
\frac{\gm_1}{\om_0} \ll 1 \; , \qquad \frac{\gm_2}{\om_0}\ll 1 \; , \qquad
\frac{|\vec d\cdot \vec E_{0i}|}{\om_0} \ll 1 \; ,
\ee
we see that $u_i$ has to be classified as a fast solution while $s_i$ and
$|u_i|^2$, as slow ones. With $s_i$ being a quasi--invariant, Eq. (111) gives
$$
u_i(t) = e^{-(i\om_0 +\gm_2)\;t}\; \left [ u_0 - i\;s_i\;\vec d \cdot
\int_0^t\vec E_{0i}(\tau)\; e^{(i\om_0 +\gm_2)\;\tau}\; d\tau\right ] \; .
$$
Let the external field $\vec E_{0i}=\vec E_{0i}(t)$ consist of two parts,
\be
\label{115}
\vec E_{0i} = \vec E_0 + \vec E_1 e^{i(\vec k\cdot \vec r_i -\om t)}
+\vec E_1^* e^{-(\vec k\cdot\vec r_i -\om t)} \; ,
\ee
one being a stationary nonresonant field $\vec E_0$, and another part is a
pair of plane waves, which are in quasiresonance with the transition
frequency,
\be
\label{116}
\frac{|\Dlt|}{\om_0} \ll 1 \; , \qquad \Dlt \equiv \om - \om_0 \; .
\ee
Then the solution of Eq. (111) writes
$$
u_i(t) = -\; \frac{s_i\;\vec d\cdot\vec E_0}{\om_0 - i\gm_2} +
\frac{s_i\;\vec d\cdot \vec E_1}{\Dlt +i\gm_2}\;
e^{i(\vec k\cdot\vec r_i - \om t)} +
$$
\be
\label{117}
+ \left ( u_0 + \frac{s_i\;\vec d\cdot\vec E_0}{\om_0 -i\gm_2} -
\frac{s_i\vec d\cdot\vec E_1}{\Dlt+ i\gm_2}\; e^{i\;\vec k\cdot\vec r_i}
\right ) \; e^{-(i\om_0 + \gm_2)\; t} \; .
\ee
Substituting this into the right--hand side of Eq. (112) and averaging over
time as
$$
\lim_{\tau\ra\infty} \; \frac{1}{\tau} \int_0^\tau f(s,t) \; dt \; ,
$$
we come to the equation
\be
\label{118}
\frac{ds_i}{dt} = -\gm_1^* \; ( s_i - \zeta^* ) \; ,
\ee
with
$$
\gm_1^* \equiv \gm_1 + 4\gm_2 \left (
\frac{|\vec d\cdot\vec E_0|^2}{\om_0^2 +\gm_0^2} +
\frac{|\vec d\cdot\vec E_1|^2}{\Dlt^2 +\gm_2^2} \right ) \; , \qquad
\zeta^* \equiv \frac{\gm_1}{\gm_1^*}\; \zeta \; .
$$
The solution to Eq. (118) is
\be
\label{119}
s_i(t) = s_0\; e^{-\gm_1^* t} +
\zeta^* \left (1 - e^{-\gm_1^* t}\right ) \; .
\ee
Calculating the correlation function
$$
\overline{u_i^*(t) u_j(t)} = s^2(t) \; \left (
\frac{|\vec d\cdot\vec E_0|^2}{\om_0^2 +\gm_2^2} +
\frac{|\vec d\cdot\vec E_1|^2}{\Dlt^2 +\gm_2^2}\;
e^{-i\;\vec k\cdot\vec r_{ij}} \right ) \; ,
$$
where, for simplicity, we set $s_i=s$, we find the incoherent and coherent
radiation intensities (84) and (85), respectively, as
$$
I_{inc}(\vec n,t) = \frac{3N}{16\pi}\; \om_0\; \gm \left |\vec n\times
\vec e_d \right |^2 \; [ 1 + s(t) ] \; ,
$$
$$
I_{coh}(\vec n,t) = \frac{3N^2}{8\pi}\; \om_0\; \gm\;  \left |\vec n\times
\vec e_d \right |^2\;  s^2(t) \; \times
$$
\be
\label{120}
\times \left [ F(k_0\;\vec n)\;
\frac{|\vec d\cdot\vec E_0|^2}{\om_0^2 +\gm_2^2} + F(k_0\;\vec n -\vec k)\;
\frac{|\vec d\cdot\vec E_1|^2}{\Dlt^2 +\gm_2^2} \right ] \; ,
\ee
where $\vec n\equiv \vec r/r$ and the form factor is
\be
\label{121}
F(\vec k) \equiv \frac{1}{N^2}\; \sum_{i\neq j}^N\;
e^{i\vec k\cdot\vec r_{ij}}
= \left | \frac{1}{N}\; \sum_{i=1}^N\;
e^{i\vec k\cdot\vec r_i}\right |^2 \; .
\ee
As is seen from expressions (120) and (121), the maxima of coherent radiation
occur in the directions satisfying the condition
\be
\label{122}
\left ( k_0\;\vec n -\vec k\right ) \cdot \vec r_i = 2\pi n_i \qquad
(n_i=0,1,2,\ldots) \; .
\ee
This corresponds either to forward scattering, when all $n_i=0$, and the
periodicity of matter is not required, or to the scattering in the Bragg
directions, for which the strict space periodicity of radiators is needed.
The enhancement of coherent radiation in the directions defined by condition
(122) is called the Borrmann effect [69,70], which for the case of
$\gm$--rays is sometimes termed the Kagan--Afanasiev effect [71,72].

The total radiation intensities (87) and (88) are
$$
I_{inc}(t) = \frac{1}{2}\; N \om_0\gm\; [ 1 + s(t) ] \; ,
$$
\be
\label{123}
I_{coh}(t) = N^2\;\vp\;\om_0\;\gm\; s^2(t) \left (
\frac{|\vec d\cdot\vec E_0|^2}{\om_0^2 +\gm_2^2} +
\frac{|\vec d\cdot\vec E_1|^2}{\Dlt^2 +\gm_2^2} \right ) \; ,
\ee
where the shape factor is
\be
\label{124}
\vp \equiv \frac{3}{8\pi} \int\left | \vec n\times \vec e_d\right |^2\;
F(k_0\;\vec n -\vec k) \; d\Omega(\vec n) \; .
\ee
The value of the latter strongly depends on the shape of the considered
sample. Thus, for pencil--like or disk--like shapes [29], one has
\begin{eqnarray}
\vp =\left\{ \begin{array}{ccc}
\frac{3\lbd}{8L} \; , & \frac{\lbd}{2\pi L} \ll 1 \; , & \frac{R}{L}\ll 1 \\
\\
\nonumber
\frac{3}{8}\left (\frac{\lbd}{\pi R}\right )^2\; , &
\frac{\lbd}{2\pi R} \ll 1 \; , & \frac{L}{R} \ll 1 \; ,
\end{array} \right.
\end{eqnarray}
where $R$ and $L$ are the radius and length of a cylindrical sample, and
$\lbd\equiv 2\pi/k, \; k\equiv|\vec k|=\om/c$.

Consider the stationary limit $t\ra\infty$, keeping in mind the situation
typical of M\"osbauer experiments, when the alternating field is weak,
\be
\label{125}
\frac{|\vec d\cdot\vec E_1|^2}{\gm_1\gm_2} \ll 1 \; ,
\ee
and let us set, for simplicity,
\be
\label{126}
\zeta = -1
\ee
which means that there is no additional pumping except through the given
field (115). Then Eq. (119) reduces to
$$
\lim_{t\ra\infty} s_i(t) = - 1  +\frac{4\gm_2}{\gm_1}\; \left (
\frac{|\vec d\cdot\vec E_0|^2}{\om_0^2} +
\frac{|\vec d\cdot\vec E_1|^2}{\Dlt^2 +\gm_2^2} \right ) \; .
$$
For the coherence coefficient, defined in Eq. (90), we get
\be
\label{127}
\lim_{t\ra\infty} C_{coh}(t) = N\; \frac{\vp\gm_1}{2\gm_2} \; .
\ee
The role of the nonresonant field $\vec E_0$ can be characterized by the
{\it switching factor} [24]
\be
\label{128}
S(E_0,t) \equiv \; \frac{I(t)}{\lim_{E_0\ra 0} I(t)}
\ee
and its stationary limit
\be
\label{129}
S(E_0) \equiv \lim_{t\ra\infty} S(E_0,t) \; .
\ee
For our case, we obtain
\be
\label{130}
S(E_0) = 1 + \frac{\Dlt^2+\gm_2^2}{\om_0^2}\;  \left |
\frac{\vec d\cdot \vec E_0}{\vec d\cdot \vec E_1}\right |^2 \; .
\ee
The switching factors (128) and (129) show how the radiation intensity is
amplified when a nonresonant field $\vec E_0$ is switched on, as compared to
the situation when $\vec E_0=0$. As is seen from expression (130), the
amplification can be quite noticeable only if
$|\vec d\cdot\vec E_0|\gg |\vec d\cdot\vec E_1|$, so that to compensate the
smallness of the parameters $|\Dlt|/\om_0$ and $\gm_2/\om_0$.

\section{M\"ossbauer Magnetic Anomaly}

Stationary fields, electric or magnetic, can be due not to external sources
but can arise in a sample as a result of phase transitions [73,74]. If an
ensemble of radiators is incorporated into matter exhibiting a phase
transition accompanied by the appearance of a constant field, the latter
may influence some radiation characteristics. An interesting example of this
kind is given by the gamma radiation of M\"ossbauer nuclei placed into
magnetic materials. This example is especially intriguing because of
long--standing controversy related to its interpretation.

There exists a number of experiments demonstrating the so--called magnetic
anomaly of the M\"ossbauer effect in materials undergoing magnetic phase
transition. This anomaly consists in an essential increase, up to $50\%$, of
the area under the M\"ossbauer spectrum below the temperature of magnetic
transition, as compared to the spectrum area in paramagnetic state above the
transition temperature. A detailed discussion of these experiments can be
found in the book [75] and review [76]. The controversy related to this
anomaly concerns the explanation of the cause of the latter.

The area of the M\"ossbauer spectrum, for M\"ossbauer nuclei in a solid
sample, is given by the integral
\be
\label{131}
A_{abs} = f_M \int_{-\infty}^{+\infty} \sgm_{abs}(\om)\; d\om \; ,
\ee
in which
\be
\label{132}
f_M = \exp(-k_0^2r_0^2)
\ee
is the M\"ossbauer factor, $k_0=\om_0/c,\; r_0$ is the mean--square deviation
of the nucleus from a lattice site,
\be
\label{133}
\sgm_{abs}(\om) =\frac{\sgm_0\Gm_{abs}^2}{(\om-\om_0)^2+\Gm_{abs}^2}
\ee
is the absorption cross--section, $\Gm_{abs}$ is the absorption half--width,
\be
\label{134}
\sgm_0 = \frac{2\pi(1+2I_1)}{k_0^2(1+2I_0)(1+\al_e)}
\ee
is the cross--section of resonant absorption, $I_0$ and $I_1$ are the nuclear
spins of the ground--state and excited levels, and $\al_e$ is the electron
conversion coefficient. After integrating Eq. (131), we have the spectrum
area
\be
\label{135}
A_{abs} = \pi f_M\sgm_0\Gm_{abs} \; .
\ee

It is important to emphasize that the M\"ossbauer anomaly, we consider here,
has been observed only in the so--called absorption geometry, when absorbing
M\"ossbauer nuclei are placed inside magnetic matter which is irradiated by
an external source. Contrary to this, in the experiments with the so--called
source geometry, when a radioactive source is incorporated into the magnetic
matter, but absorbing M\"ossbauer nuclei are outside this matter, no magnetic
anomaly has been observed [77--79]. Therefore it is clear that the considered
M\"ossbauer anomaly is directly related to the action on M\"ossbauer nuclei
of an effective magnetic field appearing below the critical point. But what
is the origin of this anomaly?

Historically, the first suggestion was to ascribe the anomaly in the
temperature behaviour of the spectrum area (135) to the influence of the
appearing magnetic order on the M\"ossbauer factor (132). A number of
citations having to do with this suggestion are listed in Refs. [75,76]. This
assumption implies that the mean--square deviation $r_0$ defining the
M\"ossbauer factor (132) is essentially influenced by arising magnetic order.
The course of reasoning is as follows. M\"ossbauer nuclei doped into a solid
are characterized by the same mean--square deviation as the particles forming
the solid sample. The latter can be described by the Hamiltonian
\be
\label{136}
\hat H_m = \sum_i \;\frac{\vec{p_i}^2}{2m} + \frac{1}{2}\; \sum_{i\neq j}\;
\Phi(R_{ij}) - \sum_{i\neq j}\; I(R_{ij})\; \vec S_i\cdot\vec S_j \; ,
\ee
in which $\Phi(R_{ij})$ is a potential of direct pair interactions while
$I(R_{ij})$ is that of exchange interactions, $\vec S_i$ is a spin operator,
and $R_{ij}\equiv |\vec R_{ij}|$, with $\vec R_{ij}=\vec R_i -\vec R_j$. The
indices of summation in Eq. (136) run as $i=1,2,\ldots,N_0$, with $N_0$
being the number of lattice sites. Introduce the deviation from a lattice
site,
\be
\label{137}
\vec u_i \equiv \vec R_i -\vec a_i \; ,
\ee
defined so that
\be
\label{138}
\vec a_i = \; <\vec R_i > \; , \qquad <\vec u_i>\; = 0 \; .
\ee
Taking into account that $|\vec u_i|$ is small as compared to the
interparticle distance, one expands the interaction potential in powers of
$u^\al_i$ up to the second order, which results in the Hamiltonian
\be
\label{139}
\hat H_m = U_0 + \hat H_p +\hat H_s +\hat H_{sp} + \hat H' \; ,
\ee
whose terms are explained below: the constant part of the lattice energy
\be
\label{140}
U_0 = \frac{1}{2}\; \sum_{i\neq j}\;  \Phi(a_{ij})\; ; \qquad
a_{ij} \equiv |\vec a_{ij}|\; , \qquad
\vec a_{ij}\equiv \vec a_i -\vec a_j\; ,
\ee
the phonon term
\be
\label{141}
\hat H_p =\sum_i \; \frac{\vec{p_i}^2}{2m} +
\frac{1}{2}\; \sum_{i\neq j}\; \sum_{\al\bt}\; \Phi_{ij}^{\al\bt}\;
u_i^\al\; u_j^\bt\; ,
\ee
in which $\Phi_{ij}^{\al\bt} \equiv \prt^2\Phi(a_{ij})/
\prt a_i^\al\prt a_j^\bt$, the spin Hamiltonian
\be
\label{142}
\hat H_s = - \sum_{i\neq j}\; I(a_{ij})\; S_{ij} \; ; \qquad
S_{ij} \equiv \vec S_i\cdot \vec S_j \; ,
\ee
the term responsible for spin--phonon interactions,
\be
\label{143}
\hat H_{sp} = - \sum_{i\neq j}\; \sum_{\al\bt}\; I_{ij}^{\al\bt}\;
u_i^\al\; u_j^\bt\; S_{ij} \; ,
\ee
where $I_{ij}^{\al\bt}= \prt^2 I(a_{ij})/\prt a_i^\al\prt a_j^\bt$, and the
term
\be
\label{144}
\hat H'= -\sum_i\; \sum_\al\;  u_i^\al\; \left ( 1 +\frac{1}{2}\;
\sum_\bt \; u_i^\bt\; \frac{\prt}{\prt a_i^\bt}\right )\;  F_i^\al
\ee
related to the striction energy, where the striction force acting on the
site $i$ is given by the components
$$
F_i^\al \equiv -\; \frac{\prt}{\prt a_i^\al}\; \sum_j\; \left [
\Phi(a_{ij}) - 2I(a_{ij}) S_{ij}\right ] \; .
$$
The correct definition of the lattice sites in Eq. (138) presupposes that
they serve as equilibrium positions for particles. This implies that the
striction energy is to be zero on average,
\be
\label{145}
<\hat H'> \; = 0 \; .
\ee
Then one invokes a kind of the semiclassical approximation
$$
< u_i^\al S_{ij}> \; = \; < u_i^\al>< S_{ij}>\; = 0 \; , \qquad
< u_i^\al u_j^\bt S_{ij}> \; = \; < u_i^\al u_j^\bt>< S_{ij}>\; ,
$$
decoupling the phonon and spin degrees of freedom, which suggests to present
the operator term in the spin--phonon interaction (143) as
\be
\label{146}
u_i^\al \; u_j^\bt\; S_{ij} = \; < u_i^\al\; u_j^\bt> \; S_{ij} +
u_i^\al\; u_j^\bt\; < S_{ij}> \; -
\; < u_i^\al\; u_j^\bt >\; < S_{ij}> \; .
\ee
Thus, the matter Hamiltonian (139) can be reduced to
\be
\label{147}
\hat H_m = \overline U_0 +\hat{\overline H}_p + \hat{\overline H}_s \; ,
\ee
with the renormalized terms
$$
\overline U_0 = U_0 + \sum_{i\neq j}\; \sum_{\al\bt}\; I_{ij}^{\al\bt}
< u_i^\al u_j^\bt> \; < S_{ij}> \; ,
$$
$$
\hat{\overline H}_p = \sum_i \; \frac{\vec{p_i}^2}{2m} +
\frac{1}{2}\; \sum_{i\neq j}\; \sum_{\al\bt}\; D_{ij}^{\al\bt}\; u_i^\al\;
u_j^\bt \; , \qquad
\hat{\overline H}_s = - \sum_{i\neq j}\; J_{ij}\; S_{ij} \; ,
$$
in which the striction energy, because of condition (145), is omitted and the
renormalized interactions are
$$
D_{ij}^{\al\bt} \equiv \Phi_{ij}^{\al\bt} -
2 I_{ij}^{\al\bt} \; < S_{ij} >\; , \qquad
J_{ij} \equiv I(a_{ij}) +\sum_{\al\bt}\; I_{ij}^{\al\bt} \;
< u_i^\al u_j^\bt > \; .
$$
The renormalized dynamical matrix $D_{ij}^{\al\bt}$ defines the effective
phonon spectrum $\om_{ks}$ through the eigenvalue problem
$$
\frac{1}{m}\; \sum_j\; \sum_\bt\; D_{ij}^{\al\bt}\;
e^{-\vec k\cdot\vec a_{ij}}\;
e_{ks}^\bt = \om_{ks}^2\; e^\al_{ks} \; ,
$$
where $\vec e_{ks}$ is a polarization vector, the index $s$ labelling
polarizations. The spectrum and polarization vectors are assumed to be even
functions of the wave vector, so that $\om_{ks} = \om_{-ks}$ and
$\vec e_{ks}=\vec e_{-ks}$. Polarization vectors form a complete orthonormal
basis with the properties
$$
\vec e_{ks}\cdot \vec e_{ks'} = \dlt_{ss'} \; , \qquad
\sum_s\; e_{ks}^\al\; e_{ks}^\bt = \dlt_{\al\bt} \; .
$$
Expanding the deviation and momentum as
$$
\vec u_i = \sum_{ks}\; \frac{\vec e_{ks}}{\sqrt{2mN_0\om_{ks}}}\; \left (
b_{ks} + b_{-ks}^\dgr \right ) \; e^{i\vec k\cdot\vec a_i} \; , \qquad
\vec p_i = -i\; \sum_{ks}\; \sqrt{ \frac{m\om_{ks}}{2N_0}} \; \vec e_{ks}\;
\left ( b_{ks} - b_{-ks}^\dgr \right )\; e^{i\vec k\cdot\vec a_i} \; ,
$$
one transforms the renormalized phonon Hamiltonian to the standard form
$$
\hat{\overline H}_p = \sum_{ks} \left ( b_{ks}^\dgr\; b_{ks} +
\frac{1}{2} \right ) \; .
$$
After this, it is straightforward to calculate the correlators
$$
< u_i^\al u_j^\bt > \; =\; \frac{\dlt_{ij}}{2N_0}\;
\sum_{ks} \; \frac{e_{ks}^\al\; e_{ks}^\bt}{m\;\om_{ks}}\; {\rm coth}\;
\frac{\om_{ks}}{2T} \; ,
$$
in which $T$ is temperature. Thus, one gets the mean--square deviation from
the equation
\be
\label{148}
r_0^2 \equiv \frac{1}{3} \; \sum_\al < u_i^\al u_i^\al > \; =
\frac{1}{6mN_0}\; \sum_{ks}\;  \frac{1}{\om_{ks}}\;
{\rm coth} \; \frac{\om_{ks}}{2T} \; .
\ee
In this way, the influence of magnetic order on the mean--square deviation
comes from its influence on the phonon spectrum.

However, the magnitude of the spin--phonon interaction, renormalizing the
dynamical matrix, is rather small, as compared to the magnitude of direct
interactions [76], so that $|I_{ij}^{\al\bt}/\Phi_{ij}^{\al\bt}|\sim 10^{-3}$.
Hence, magnetic order cannot influence much phonon frequencies, as well as the
sound velocity
\be
\label{149}
c_s \equiv \lim_{k\ra 0}\; \frac{\om_{ks}}{k} = - \lim_{k\ra 0}\;
\sum_j \; \sum_{\al\bt}\;  D_{ij}^{\al\bt} \;
\frac{(\vec k\cdot\vec a_{ij})^2}{2mk^2} \; e^\al_{ks}\; e^\bt_{ks} \; .
\ee
This conclusion is in agreement with all known experiments where phonon
characteristics have been examined by means of neutron scattering,
sound--velocity measurements, elastic and thermal investigations. The onset
of magnetic order can change the M\"ossbauer factor not more than by
$1\%$, which cannot explain the observed M\"ossbauer anomaly of the spectrum
area (135).

Another explanation was advanced by Babikova et al. [78], supposing that
magnetic order can influence the electron conversion coefficient $\al_e$ in
the cross--section (134). A noticeable decrease of the conversion coefficient
could lead to the increase of the cross--section (134), and, consequently,
to the increase of the spectrum area (135). The decrease of the conversion
coefficient could be due to the suppression of the conversion channel in
favour of the $\gamma$--radiation channel whose weight could be increased by
the enhancement of the $\gm$--radiation caused by the arising magnetic order
[80].

To estimate the influence of an effective magnetic field, appearing in
magnets, on the radiation intensity of M\"ossbauer nuclei, we have to consider
the switching factor (130) that in our case, takes the form
$$
S(H_0) =  1 +\frac{\gm_2^2}{\om_0^2}\;  \left |
\frac{\vec\mu\cdot\vec H_0}{\vec\mu\cdot\vec H_1}\right |^2 \; .
$$
For the characteristic M\"ossbauer nucleus $^{57}Fe$, we have
$\om_0= 1.44\times 10^4$eV and $\gm_2=\gm_1=0.67\times 10^{-8}$eV, which can
be transformed to the frequency units as $\om_0\sim 10^{19}$s$^{-1}$ and
$\gm_2\sim\gm_1\sim 10^7$s$^{-1}$. The corresponding wavelength is
$\lbd\sim 10^{-8}$cm. Let us take for the effective magnetic field
$H_0\sim 10^5$G and for the alternating source field $H_1\sim 10^{-5}$G.
The transition magnetic dipole $\mu_0\sim 0.1\mu_n$, where $\mu_n$ is the
nuclear magneton, hence $\mu_0\sim 10^{-13}$eV/G. This gives
$\mu_0H_0\sim 10^7$s$^{-1}$ and $\mu_0H_1\sim 10^{-3}$s$^{-1}$. From
here we obtain $\gm_2^2H_0^2/\om_0^2H_1^2\sim 10^{-4}$, which tells us that
the switching factor $S(H_0)$ changes too little. Therefore, although the
arising magnetic order does enhance the radiation of M\"ossbauer nuclei,
this enhancement is not sufficient for causing such a drastic increase of the
spectrum area.

The last quantity that could be blamed to be responsible for the M\"ossbauer
magnetic anomaly is the absorption width $\Gm_{abs}$. The latter can be
presented as the sum
\be
\label{150}
\Gm_{abs} = \gm_2 +\gm_2^*
\ee
of the homogeneous line width $\gm_2$ and of the inhomogeneous  line width
$\gm_2^*$. The inhomogeneous width can be due to the variation of local
magnetic fields resulting in the random shift of the M\"ossbauer transition
frequency [81]. Returning to Section 3, we see that, really, an external
magnetic field shifts the transition frequency as
$\om_0 +\left (\vec\mu_{22}-\vec\mu_{11}\right )\cdot\vec H_0$. Therefore,
the inhomogeneous width can be of order $\gm_2^*\sim\left (\vec\mu_{22}-
\vec\mu_{11}\right )\cdot\vec H_0$ or $\gm_2^*\sim\mu_0H_0$. From here,
$\gm_2^*\sim 10^7$s$^{-1}$, that is, $\gm_2^*\sim\gm_2$. In this way, the
anomalous increase of the M\"ossbauer spectrum area (135) below the magnetic
transition temperature can be explained by the increase of the absorption
width (150) caused by the increasing inhomogeneous width
$\gm_2^*\sim\mu_0 H_0$.

\section{Problem of Pattern Selection}

Nonequilibrium cooperative phenomena are often described by nonlinear
differential or integro--differential equations in partial derivatives. The
solutions to such equations are in many cases nonuniform in space exhibiting
the formation of different spatial structures. It happens that a given set
of equations possesses several solutions corresponding to different spatial
patterns [13]. In such a case, the question arises which of these solutions,
and respectively patterns, to prefer? The problem of pattern selection has
no general solution [13]. A possible way of selecting spatial structures,
by minimizing the average energy, was delineated in subsection 2.5. Here we
advance another method of pattern selection.

Assume that the considered differential equations in partial derivatives can
be reduced to a $d$--dimensional system of ordinary equations; the
dimensionality $d$ may equal infinity. Suppose also that admissible patterns
are parametrized by a multiparameter $\bt$. Let the state of the dynamical
system be defined by the set
\be
\label{151}
y(t) =\{ y_i(t)=y_i(\bt,t)| \; i=1,2,\ldots, d\}
\ee
of solutions to the system of differential equations
\be
\label{152}
\frac{d}{dt}\; y(t) = v(y,t) \; .
\ee
For different parameters $\bt$ there are different sets (151) corresponding
to different spatial structures. All admissible values of $\bt$ form a
manifold ${\cal B}=\{\bt\}$. Each particular value of $\bt$ can be considered
as a realization of the random variable from the manifold ${\cal B}$. The
classification of the states (151) can be done by defining a probability
measure on ${\cal B}$.

To introduce the probability distribution $p(\bt,t)$ of patterns at time $t$,
we resort to the ideas of statistical mechanics [82], where a probability
$p$ can be connected with entropy $S$ by the relation $p\sim e^{-S}$. The
entropy at time $t$ may be expressed as
\be
\label{153}
S(t) \equiv\ln|\Dlt\Phi(t)|
\ee
through the elementary phase volume
\be
\label{154}
\Dlt\Phi(t) \equiv \prod_i\dlt\; y_i(t) \; .
\ee
Let us count the entropy from its initial value $S(0)$, thus, considering
the entropy variation
\be
\label{155}
\Dlt S(t) \equiv S(t) - S(0) \; .
\ee
Then the probability distribution $p\sim e^{-\Dlt S}$, normalized by the
condition
$$
\int p(\bt,t)\; d\bt = 1
$$
takes the form
\be
\label{156}
p(\bt,t) = \frac{e^{-\Dlt S(\bt,t)}}{Z(t)} \; ,
\ee
where the normalization factor is
$$
Z(t) = \int e^{- \Dlt S(\bt,t)}\; d\bt \; .
$$
The entropy variation (155) writes
\be
\label{157}
\Dlt S(t) = \ln \left | \frac{\Dlt\Phi(t)}{\Dlt\Phi(0)}\right | \; ,
\ee
where the dependence on $\bt$, for brevity, is omitted. Define the multiplier
matrix [83]
\be
\label{158}
M(t) = [M_{ij}(t)] \; , \qquad M_{ij}(t) \equiv
\frac{\dlt y_i(t)}{\dlt y_j(0)} \; ,
\ee
for which at the initial time one has
\be
\label{159}
M_{ij}(0) \equiv
\frac{\dlt y_i(0)}{\dlt y_j(0)} = \dlt_{ij} \; .
\ee

The variation of the state (151) gives
\be
\label{160}
\dlt y(t) = M(t)\; \dlt y(0) \; ,
\ee
which yields for the elementary phase volume (154)
$$
\Dlt\Phi(t) =\prod_i\; \sum_{j}\; M_{ij}(t)\; \dlt y_j(0) \; .
$$
Hence, the entropy variation (157) is
$$
\Dlt S(t) = \ln\left | \prod_i\; \sum_j\; M_{ij} (t)\; M_{ji}(0)
\right | \; .
$$
With condition (159), this results in
\be
\label{161}
\Dlt S(t) = \ln \left | \prod_i \; M_{ii}(t) \right |= \sum_i\;
\ln| M_{ii}(t)| \; .
\ee
Taking the variational derivative of equation (152), we get the equation
\be
\label{162}
\frac{d}{dt} M(t) = J(y,t)\; M(t)
\ee
for the multiplier matrix (158), where
\be
\label{163}
J(y,t) = [ J_{ij}(y,t) ]\; , \qquad
J_{ij}(y,t) \equiv \frac{\dlt v_i(y,t)}{\dlt y_j(t)} \; ,
\ee
is the Jacobian matrix. Substituting the entropy variation (161) into
Eq. (156), we get
\be
\label{164}
p(\bt,t) = \frac{\prod_i | M_{ii}(\bt,t)|^{-1}}{Z(t)} \; ,
\ee
with
$$
Z(t) = \int \prod_i | M_{ii}(\bt,t)|^{-1} \; d\bt \; .
$$
Expression (164) defines the probability distribution of patterns labelled by
a multiparameter $\bt$. This expression naturally connects the notion of
probability and the notion of stability. Really, the multipliers are smaller
by modulus for more stable solutions and, respectively, patterns, for which
the probability is higher.

Another form of the distribution (164) can be derived as follows. Introduce
the matrix
\be
\label{165}
L(t) = [ L_{ij}(t) ] \; , \qquad L_{ij}(t) \equiv \ln| M_{ij}(t)| \; .
\ee
Then the entropy variation (161) becomes
\be
\label{166}
\Dlt S(t) = {\rm Tr}\; L(t) \; .
\ee
Since the trace of a matrix does not depend on its representations, we may
perform intermediate transformations of Eq. (166) using one particular
representation and returning at the end to the form independent of
representations. To this end, let us consider a representation when the
multiplier matrix is diagonal. Because of Eq. (162) with the initial
condition (159), the matrix $M$ is diagonal if and only if the Jacobian
matrix is also diagonal. Then from the evolution equation (162) it follows
that
$$
M_{ii}(t) = \exp\left\{ \int_0^t J_{ii}(y(t'),t')\; dt'\right\} \; .
$$
Hence
$$
L_{ii}(t) = \int_0^t \Lbd_i(t') \; dt'\; , \qquad
\Lbd_i(t) \equiv {\rm Re}\; J_{ii}(t) \; ,
$$
from where
$$
{\rm Tr}\; L(t) = \int_0^t \Lbd(t')\; dt'\; , \qquad
\Lbd(t) \equiv \sum_i \; \Lbd_i(t) \; .
$$
We assume that the state (151) is formed of real functions, so that the
velocity field in the evolution equation (152) is also real. Then the
eigenvalues of the Jacobian matrix (163) are either real or, if complex,
come in complex conjugate pairs. Therefore
$$
\sum_i\; {\rm Re}\; J_{ii}(y,t) = \sum_i\; J_{ii}(y,t) = 
{\rm Tr}\; J(y,t) \; .
$$
For the entropy variation (166) we obtain
\be
\label{167}
\Dlt S(t) = \int_0^t \Lbd(t')\; dt'\; ,
\ee
where
\be
\label{168}
\Lbd(t) ={\rm Tr} \; J(y,t)
\ee
is called [84] the {\it contraction rate}. The latter is given by the form
independent of representations of the Jacobian matrix (163). With the entropy
variance (167), the probability distribution (156) becomes
\be
\label{169}
p(\bt,t) =\frac{1}{Z(t)}\;  \exp\left\{ -\int_0^t \Lbd(\bt,t')\;
dt'\right\} \; ,
\ee
where the contraction rate is defined in Eq. (168) and
$$
Z(t) = \int\exp\left\{ - \int_0^t\Lbd(\bt,t')\; dt'\right \}\; d\bt \; .
$$
The {\it most probable pattern} at a time $t$ corresponds to the maximum
of the distribution (169),
\be
\label{170}
{\rm abs}\; \max_\bt\;  p(\bt,t) \ra \bt(t) \; .
\ee
One may also define the {\it average pattern} at $t$ as corresponding to
$$
\overline{\bt}(t) \equiv \int \bt\; p(\bt,t)\; d\bt \; .
$$
The most probable and average patterns, in general, do not coincide, although
this may happen, especially with increasing time. To illustrate the latter,
consider a particular case when the contraction rate $\Lbd(\bt,t)=\Lbd(\bt)$
does not depend on time. Then, as $t\ra\infty$, we have
$$
Z(t) = \int e^{-\Lbd(\bt)\; t} \; d\bt \simeq
\sqrt{\frac{2\pi}{\Lbd''(\bt_0)t}} \; 
\exp\left\{ - \Lbd(\bt_0)\; t\right\} \; ,
$$
where $\bt_0$ is the point of the minimum of $\Lbd(\bt)$, so that
$$
\frac{d}{d\bt}\; \Lbd(\bt) = 0 \; , \qquad
\Lbd''(\bt) \equiv \frac{d^2}{d\bt^2}\; \Lbd(\bt) > 0 
\qquad (\bt=\bt_0)\; .
$$
In the distribution
$$
p(\bt,t) \simeq \; \sqrt{\frac{\Lbd''(\bt_0)}{2\pi}\; t} \; \exp\left\{ -
[ \Lbd(\bt) - \Lbd(\bt_0) ] \; t \right\}
$$
one may expand $\Lbd(\bt)$ near $\bt=\bt_0$, which gives
$$
p(\bt,t) \simeq \frac{1}{\sqrt{2\pi}\sgm(t)} \; \exp\left\{ -
\frac{(\bt-\bt_0)^2}{2\sgm^2(t)}\right\} \; , \qquad
\sgm(t) \equiv \frac{1}{\sqrt{\Lbd''(\bt_0)\; t}}\; .
$$
From here one finds
$$
\lim_{t\ra\infty}\; p(\bt,t) = \dlt(\bt-\bt_0) \; .
$$

In this way, if differential equations describing a nonequilibrium process
have several solutions corresponding to different spatial patterns, the latter
can be characterized by the probability distribution (169), with the
contraction rate (168). In the case when the multiplier matrix (158) can be
calculated, one may use the expression (164) of the probability distribution.
If all patterns correspond to stable solutions, it is sufficient to analyse
only the beginning of the process of pattern formation. Then for the entropy
variation (167) we may write
$$
\Dlt S(\bt,t) \simeq \Lbd(\bt,0)\; t \qquad (t\ra 0) \; .
$$
Consequently, the most probable pattern, defined by the maximum of the
probability distribution (169), that is, by the minimum of the entropy
variation (167), is now characterized by the minimum of the contraction rate
$\Lbd(\bt,0)$ at the initial time.

\section{Turbulent Photon Filamentation}

Spatial structures can appear in radiating systems if the radiation wavelength
is much shorter than the system characteristic sizes [13]. For instance,
electric field in laser cavities can exhibit a state which bears some analogy
with a superfluid vortex [85]. The Maxwell--Bloch equations for slowly
varying field amplitudes have been shown to be analogous to hydrodynamic
equations for compressible viscous fluid [86]. The Fresnel number for optical
systems plays the role similar to the Reynolds number for fluids. In the
same way as when increasing the Reynolds number, the fluid becomes turbulent,
there can appear optical turbulence when increasing the  Fresnel number.

Spatial structures emerge from an initially homogeneous state with a break
of space--translational symmetry. For small Fresnel numbers $F\leq 5$, such
structures correspond to the empty--cavity Gauss--Laguerre modes imposed
by the cavity geometry. These transverse structures can be described by
expanding fields over the modal Gauss--Laguerre functions [87--92], which
results in reasonable agreement with experiments for CO$_2$ and Na$_2$
lasers.  For large Fresnel numbers $F>10$, the appearing structures are
very different from those associated with empty--cavity modes. The modal
expansion is no longer relevant at large $F$, and the boundary conditions
have little or no importance. The laser medium looks like divided in a
large amount of parallel independently oscillating uncorrelated filaments
[93--100] the number of filaments being proportional to $F$, contrary to
the case of small Fresnel numbers when the number of bright spots is
proportional to $F^2$. This filamentation was observed in Dye and CO$_2$
lasers, as well as in other resonance media, even without resonators
[101--105]. The same type of patterns arises in active nonlinear media,
such as photorefractive Bi$_{12}$SiO$_{20}$ crystal pumped by a laser
[106--109]. In the latter media there are also two types of pattern
formation: for small Fresnel numbers, the symmetry is imposed through the
boundary, while for large Fresnel numbers, the symmetry is imposed by the
bulk parameters. In the case of large $F$, there occurs a kind of
self--organization with spontaneous spatial symmetry breaking [110]. It is
possible to easily notice a qualitative transition in the behaviour of
photorefractive media as well as in that of lasers: In low--$F$ regime there
are a few modes of regular arrangement of bright spots corresponding to the
peaks of the Gauss--Laguerre functions in cylindrical geometry, the number
of modes being proportional to $F^2$. And in the high--$F$ regime there are
many modes spatially uncorrelated with each other, which is typical for
spatiotemporal chaos, the number of the chaotic filaments being proportional
to $F$. Short--range spatial correlation is characteristic for turbulence,
this is why one calls the similar phenomenon in optics the optical 
turbulence.

The theory of self--organized photon filamentation in high--Fresnel--number
resonant media was suggested in Refs. [33,111-116], where the consideration
was based on simplified models and only the stationary regime was analysed.
The choice of filament radii was done by means of the variational principle,
as is described in subsection 2.5. Here we present a more general and
elaborate theory based on the evolution equations (68) to (70), which
includes the description of temporal behaviour, and for defining the
characteristics of filaments we employ the method of pattern selection
developed in Sec. 9.

First, it is convenient to pass in Eqs. (68) to (70) to continuous
representation replacing the sums by integrals according to the rule
$$
\sum_{i=1}^N = \int\rho(\vec r)\; d\vec r \; ,
$$
where $\rho(\vec r)$ is the spatial density of radiators. Wishing to return
to the localized representation, one makes the replacement
$\rho(\vec r)= \sum_{i=1}^N\dlt(\vec r-\vec r_i)$. In the case when the
structure of matter is of no importance, it can be treated as uniform on
average setting $\rho(\vec r)=\rho\equiv N/V$. Cooperative optical phenomena
are often considered in this representation of uniform medium [117]. Let us
stress that the uniformity of matter in no case requires the uniformity of
fields or polarization. The solutions to Eqs. (68) to (70) can correspond
to highly nonuniform structures.

Introduce the notation
\be
\label{171}
f(\vec r,t) \equiv f_0(\vec r,t) + f_{rad}(\vec r,t)
\ee
for an effective field acting on a radiator with the transition dipole
$\vec d$. This field consists of the term
\be
f_0(\vec r,t) \equiv -i\; \vec d\cdot \vec E_0(\vec r,t)
\ee
due to an external electric field and of the term
\be
\label{173}
f_{rad}(\vec r,t) \equiv k_0\; <\vec d\cdot\vec A_{rad}(\vec r,t)>
\ee
responsible for the action of other radiators. Taking into account Eq. (65),
we have
\be
\label{174}
f_{rad}(\vec r,t) = -\; \frac{3}{4}\; i\gm\rho 
\int\left [\vp(\vec r-\vec r\;')\;  u(\vec r\;',t) - 
\vec{e_d}^2\; \vp^*(\vec r-\vec r\;')\;
u^*(\vec r\;',t)\right ]\; d\vec r\;'\; ,
\ee
where the continuous representation is used, and
$$
\vp(\vec r) \equiv \frac{e^{ik_0|\vec r|}}{k_0|\vec r|} \; , \qquad
\gm\equiv \frac{4}{3}\; k_0^3\; d_0^2\; .
$$
Then Eqs. (68) to (70) acquire the form
$$
\frac{du}{dt} = - (i\om_0+\gm_2) u + sf \; , \qquad
\frac{ds}{dt} = - 2( u^*f + f^*u) - \gm_1(s-\zeta) \; ,
$$
\be
\label{175}
\frac{d|u|^2}{dt} = - 2\gm_2 |u|^2 + s( u^*f + f^* u) \; .
\ee
Notice that from the latter two equations one has
$$
\frac{d}{dt} \left ( s^2 + 4|u|^2\right ) = -2\gm_1 s(s-\zeta) -
8\gm_2|u|^2 \; .
$$

We consider a sample of the cylindrical shape typical of lasers. The seed
laser field defining the cylinder axis is given by the sum of two running
waves,
\be
\label{176}
\vec E_0 (\vec r,t) = \vec E_1\; e^{i(kz-\om t)} +
\vec E_1^* \; e^{-i(kz-\om t)} \; ,
\ee
which selects a longitudinal mode. The radius, $R$, and length, $L$, of the
cylinder are such that the following inequalities are valid:
\be
\label{177}
\frac{a}{\lbd} \ll 1\; , \qquad \frac{\lbd}{R} \ll 1 \; , \qquad
\frac{R}{L}\ll 1 \; ,
\ee
where $a$ is the mean distance between radiators and $\lbd$, wavelength.
There are also the standard small parameters
\be
\label{178}
\frac{\gm_1}{\om_0} \ll 1 \; , \qquad \frac{\gm_2}{\om_0} \ll 1\; , \qquad
\frac{|\Dlt|}{\om_0} \ll 1\; ,
\ee
with $\Dlt\equiv \om-\om_0$ being detuning.

The solutions to Eqs. (175) are not necessarily uniform in the whole volume
$V=\pi R^2L$ of the sample, but may have noticeable values only inside
narrow regions of filamentary form, while being almost zero outside these
filaments. Consider one such filament, and let us surround it by a cylinder
of radius $b$ so  that the magnitude of solutions is an order smaller at the
surface of this enveloping cylinder than at its axis. If the profile of a
filament is close to the Gaussian $\exp(-r^2/2r_f^2)$, with $r_f$ being the
filament radius, then
\be
\label{179}
b= \sqrt{2\ln 10}\; r_f \; .
\ee
In what follows we assume this relation between the radius $b$ of an
enveloping cylinder and the radius $r_f$ of a filament.

Suppose that there are $N_f$ filaments in the volume of the sample, the axis
of each filament being centered at a point $\{ x_n,y_n\}$, with
$n=1,2,\ldots,N_f$. Let us present the solutions to Eqs. (175) as expansions
over enveloping cylinders,
\be
\label{180}
u(\vec r,t) = \sum_{n=1}^{N_f}\; u_n(\vec r,t)\; \Theta_n(x,y)\; 
e^{ikz} \; , \qquad
s(\vec r,t) = \sum_{n=1}^{N_f}\; s_n(\vec r,t)\; \Theta_n(x,y) \; ,
\ee
where
$$
\Theta_n(x,y) \equiv \Theta\left ( b -\sqrt{(x-x_n)^2 + (y-y_n)^2}
\right )
$$
is a unit--step function. The filaments are located randomly in the
cross--section of the sample, but so that their enveloping cylinders do not
intersect with each other. The interaction between filaments is small, which
follows from Eq. (174). This is why they do not form a regular lattice but
are distributed randomly.

The function $\vp(\vec r)$ in Eq. (174) oscillates at the distance $\lbd$,
and the solutions $u_n$ and $s_n$ essentially change in the radial direction
in the interval $b$. Assuming that
\be
\label{181}
\frac{\lbd}{b}\ll 1 \; ,
\ee
we may say that, in the radial direction, the function $\vp(\vec r)$ is
fastly varying in space, as compared to the slow variation of $u_n$ and
$s_n$. For the latter, we define the averages
\be
\label{182}
u(t) \equiv \frac{1}{V_n}\; \int_{{\bf V}_n} u_n(\vec r,t)\; d\vec r\; ,
\qquad
s(t) \equiv \frac{1}{V_n}\; \int_{{\bf V}_n} s_n(\vec r,t) \; d\vec r
\ee
over the corresponding enveloping cylinder of the volume $V_n\equiv \pi b^2L$,
where in the left--hand side of Eq. (182) we, for short, do not write the
index $n$.

The seed field (176) is needed mainly for selecting a longitudinal mode with
cylindrical symmetry, but the amplitude of this field is small, so that
\be
\label{183}
\frac{|\vec d\cdot\vec E_1|}{\gm_2} \ll 1 \; .
\ee
The excitation of radiators is accomplished by means of pumping characterized
by the pumping parameter $\zeta$ in Eqs. (175).

Defining the effective coupling parameters
\be
\label{184}
g\equiv \frac{3\gm\rho}{4\gm_2V_n}\;  \int_{{\bf V}_n}
\frac{\sin [k_0|\vec r-\vec r\;'| - k(z-z') ]}{k_0|\vec r-\vec r\;'|}\;
d\vec r \; d\vec r\;'\; ,
\ee
\be
\label{185}
g'\equiv \frac{3\gm\rho}{4\gm_2V_n} \; \int_{{\bf V}_n}
\frac{\cos [k_0|\vec r-\vec r\;'| - k(z-z') ]}{k_0|\vec r-\vec r\;'|}\;
d\vec r \; d\vec r\;'\; ,
\ee
and the collective frequency and width, respectively,
\be
\label{186}
\Om\equiv \om_0 + g'\gm_2 s \; , \qquad \Gm \equiv\gm_2 ( 1 -gs) \; ,
\ee
for functions (182) we obtain the equations
$$
\frac{du}{dt} = -(i\Om +\Gm) u - is\vec d\cdot\vec E_1 e^{-i\om t} \; ,
$$
\be
\label{187}
\frac{ds}{dt} = - 4g\gm_2|u|^2 - \gm_1( s - \zeta) - 4{\rm Im}\;
\left (u^*\; \vec d\cdot\vec E_1\; e^{-i\om t}\right ) \; ,
\ee
$$
\frac{d|u|^2}{dt} = -2\Gm|u|^2 + 2s\; {\rm Im}\; \left ( u^*\; \vec d\cdot
\vec E_1\; e^{-i\om t} \right ) \; .
$$
Because of the inequalities (178) and (183), the solution $u$ in Eqs. (187)
is fast, while $s$ and $|u|^2$ are slow in time. Using the scale separation
approach, we find
\be
\label{188}
u(t) = u_0\;  e^{-(i\Om +\Gm)t} +
\frac{s\; \vec d\cdot\vec E_1}{\om-\Om+i\Gm}\; \left [
e^{-i\om t} - e^{-(i\Om +\Gm)t} \right ] \; .
\ee

Introduce the parameter
\be
\label{189}
\al \equiv \lim_{\tau\ra\infty} \; \frac{{\rm Im}}{\tau\Gm s}\;
\int_0^\tau u^*(t) \; \vec d\cdot \vec E_1\;  e^{-i\om t}\; dt\; ,
\ee
characterizing the coupling of radiators with the seed field. This, with
Eq. (188), gives
\be
\label{190}
\al = \frac{|\vec d\cdot\vec E_1|^2}{(\om-\Om)^2+\Gm^2}\; .
\ee
The latter, according to inequality (183), is small,
\be
\label{191}
|\al|\ll 1 \; .
\ee
Finally, defining the function
\be
\label{192}
w\equiv |u|^2 -\al s^2 \; ,
\ee
we obtain the equations
\be
\label{193}
\frac{ds}{dt} = - 4g\gm_2 w - \gm_1(s-\zeta) \; , \qquad
\frac{dw}{dt} = - 2\gm_2 ( 1 -gs) \; w \; .
\ee

The behaviour of solutions to Eqs. (193) essentially depends on the values
of the coupling parameters (184) and (185). To evaluate the latter, we may
notice that their integrands diminish and fastly oscillate at the distance
of the wavelength $\lbd$. If condition (181) holds, we may neglect boundary
effects in the integrals (184) and (185) writing approximately
$$
\int_{{\bf V}_n} f(\vec r -\vec r\;')\; d\vec r\; d\vec r\;'  \cong V_n
\int_{{\bf V}_n} f(\vec r)\; d\vec r\; .
$$
Then parameter (184) reduces to
$$
g=\frac{3\pi\gm\rho}{2\gm_2}\; \int_0^b r\; dr\; 
\int_{-L/2}^{L/2}\; \frac{\sin(k_0\sqrt{r^2+z^2}-kz)}{k_0\sqrt{r^2+z^2}}
\; dz\; ,
$$
where $r$ is the radial variable. Because of the quasiresonance condition
$|\Dlt|\ll\om_0$, we have $k_0\simeq k$. With the change of the variable
$x\equiv k(\sqrt{r^2+z^2}-z)$, we get
$$
g= \frac{3\pi\gm\rho}{2\gm_2k}\; \int_0^b r\; dr \;
\int_{kr^2/L}^{kL} \; \frac{\sin x}{x}\; dx \; .
$$
In this expression, one can replace $kL\ra\infty$, thus obtaining
$$
g= \frac{3\pi\gm\rho}{2\gm_2 k} \int_0^b \left [ \frac{\pi}{2} -
{\rm Si}\left (\frac{kr^2}{L}\right ) \right ] r\; dr \; ,
$$
where the integral sine appears,
$$
{\rm Si}(x) \equiv \int_0^x \frac{\sin t}{t}\; dt = \frac{\pi}{2} +
{\rm si}(x)\; , \qquad {\rm si}(x) \equiv \int_{\infty}^x
\frac{\sin t}{t} \; dt \; .
$$
Introducing the dimensionless quantity
\be
\label{194}
\bt \equiv\frac{kb^2}{L} = \frac{2\pi b^2}{\lbd L} \; ,
\ee
we come to the coupling parameter
\be
\label{195}
g = g(\bt) = \frac{3\pi\gm\rho L}{4\gm_2 k^2} \int_0^\bt
\left [ \frac{\pi}{2} - {\rm Si}(x) \right ] \; dx \; .
\ee
This can be integrated explicitly by means of the property
$$
\int {\rm Si}(x)\; dx = x\; {\rm Si}(x) +\cos x \; ,
$$
which results in
\be
\label{196}
g(\bt) = \frac{3\pi\gm\rho L}{4\gm_2 k^2}\;  \left\{ \bt\; \left [
\frac{\pi}{2} -{\rm Si}(\bt)\right ] + 1 -\cos \bt\right \} \; .
\ee

For the coupling parameter (185), one similarly finds
\be
\label{197}
g'= g'(\bt) = -\; \frac{3\pi\gm\rho L}{4\gm_2 k^2}\;  \int_0^\bt
{\rm Ci}(x) \; dx \; ,
\ee
where the integral cosine occurs,
$$
{\rm Ci}(x) \equiv \int_\infty^x \; \frac{\cos t}{t} \; dt \; .
$$
Integrating
$$
\int {\rm Ci}(x) \; dx = x\; {\rm Ci}(x) -\sin x \; ,
$$ we finally get
\be
\label{198}
g'(\bt) = \; \frac{3\pi\gm\rho L}{4\gm_2 k^2}\;  \left [\sin\bt -
\bt\; {\rm Ci}(\bt)\right ] \; .
\ee
To better understand the properties of the coupling parameters, we consider
two limiting cases. When $x\ll 1$, then
$$
{\rm Si}(x) \simeq x -\frac{x^3}{18} \; , \qquad
{\rm Ci}(x) \simeq \gm_E +\ln x -\frac{x^2}{4} \; ,
$$
where $\gm_E=0.577216$ being the Euler constant. From here
$$
g(x) \simeq \; \frac{3\pi\gm\rho L}{4\gm_2 k^2}\; \left ( \frac{\pi}{2}\; 
x - \frac{1}{2}\;  x^2 \right ) \; , \qquad
g'(x) \simeq \; \frac{3\pi\gm\rho L}{4\gm_2 k^2}\; x\; |\ln x| \; .
$$
In the opposite case, when $x\gg 1$, using
$$
{\rm Si}(x) \simeq \frac{\pi}{2} - \frac{\cos x}{x} - \frac{\sin x}{x^2} \; ,
\qquad {\rm Ci}(x) \simeq \frac{\sin x}{x} -\frac{\cos x}{x^2}\; ,
$$
we find
$$
g(x) \simeq\;  \frac{3\pi\gm\rho L}{4\gm_2 k^2} \; \left ( 1 + 
\frac{\sin x}{x} \right ) \; , \qquad
g'(x) \simeq \;\frac{3\pi\gm\rho L}{4\gm_2 k^2} \; \left ( 
\frac{\cos x}{x} \right ) \; .
$$
These asymptotic expressions help to analyse the dependence of the coupling
parameters on the variable (194) changing in the interval
\be
\label{199}
0 < \bt \leq 2 F \qquad \left ( F\equiv \frac{\pi R^2}{\lbd L}\right ) \; .
\ee

The stability analysis of Eqs. (193), similarly to that given in Ref. [58],
shows that, for $g\zeta <1$, the solutions tend to the stationary stable
point $s_1^*=\zeta,\; w_1^*= 0$, while for $g\zeta >1$, the stable fixed
point is
$$
s_2^* =\frac{1}{g} \; , \qquad w_2^* =\frac{\gm_1\; 
(g\zeta -1)}{4g^2\gm_2}\; .
$$
In this way, for all $\bt$ from the interval (199), except the sole case when
$g\zeta=1$, there exists a stable fixed point, that is, almost all solutions
are stable, independently of the value of $\bt$. Following the method of
pattern selection from Sec. 9, we can equip the solutions labelled by $\bt$
with the probabilistic weights (169). The most probable, among all stable
solutions, is that providing the minimum of the initial contraction rate,
which for this case is
\be
\label{200}
\Lbd(\bt,0) = -\gm_1 - 2\gm_2\; ( 1 - gs_0) \; .
\ee
The minimum of this rate requires that
\be
\label{201}
\frac{dg}{d\bt} = 0\; , \qquad s_0 \; \frac{d^2g}{d\bt^2} > 0 \; .
\ee
For $s_0>0$, one needs the minimum of $g$, which gives $\bt=4.9$. From
Eq. (194), one has $b=0.88\sqrt{\lbd L}$. And the relation (179) yields
\be
\label{202}
r_f = 0.41\sqrt{\lbd L} \; \qquad (s_0>0) \; .
\ee
When $s_0<0$, conditions (201) imply the maximum of $g$, for which
$\bt=1.92,\; b=0.55\sqrt{\lbd L}$, and the filament radius is
\be
\label{203}
r_f=0.26\sqrt{\lbd L} \qquad (s_0 <0) \; .
\ee
This is practically the same value as found  for the filaments radius in
Refs. [33,111--115] by using the variational principle of subsection 2.5.
When the system of radiators is not inverted at the initial time and becomes
excited by means of a pulse characterized by the pumping parameter $\zeta$,
one has to consider the filament radius (203) as corresponding to the most
probable pattern. The number of filaments can be defined from the
normalization condition
\be
\label{204}
\frac{1}{V} \int s(\vec r,t) \; d\vec r =\zeta \; ,
\ee
assuming that the population difference equals $+1$ inside each filament
of radius $r_f$ and $-1$ outside of the filaments. Then the number of 
filaments is 
\be
\label{205}
N_f =\;\frac{1}{2}\; (1+\zeta) \left (\frac{R}{r_f}\right )^2 \; .
\ee
The most probable filament radius (203) and the number of filaments (205)
are in good agreement with the values observed in experiments [93--99,101--105].
The considered phenomenon of filamentation can be termed turbulent since the
filaments are chaotically distributed in space and for sufficiently strong
pumping, when $g\zeta > 1 +\gm_1/8\gm_2$, each filament is aperiodically
flashing in time. The turbulent photon filamentation is a self--organized
phenomenon due to the bulk properties of interacting radiators. It practically
does not depend on boundary conditions and exists in both types of lasers, the
resonator--cavity lasers, such as CO$_2$ and Dye lasers [93--99], as well as
in the resonatorless discharge--tube lasers, such as lasers on Ne, Tl,
Pb, N$_2$, and N$_2^+$ vapors [101--105]. The turbulent filamentation is also
principally nonlinear phenomenon. Thus, in low--Fresnel--number lasers
($F\leq 5$) the number of light spots is proportional to $F^2$. The same
dependence of the number of coherent rays on $F$ is typical of the initial
linearized stage of superfluorescence [118]. However, for high--$F$ lasers
($F\gg 10$) the number of filaments is proportional to $F$, which is in
agreement with formula (205) giving $N_f\sim F$.

\section{Superradiant Spin Relaxation}

When the initial state of a spin system is strongly nonequilibrium, different
kinds of spin relaxation can occur. If there are no transverse external
fields acting on spins, they relax to an equilibrium state by an exponential
law with a longitudinal relaxation time $T_1$. When the motion of spins is
triggered by a transverse magnetic field, the relaxation is again exponential
but with a transverse relaxation time $T_2$ that is usually much shorter than
$T_1$. A rather special relaxation regime arises, if the spin system is
coupled to a resonator. This can be done by inserting the sample into a
coil connected with a resonance electric circuit. Because of the action of
resonator feedback field, the motion of spins can become highly coherent
resulting in their ultrafast relaxation during a characteristic collective
relaxation time much shorter than $T_2$ [119]. This latter type of collective
spin relaxation from a strongly nonequilibrium state in the presence of
coupling with a resonator is the most difficult to realize experimentally and
to describe theoretically. Experimental difficulties have been overcome in
a series of observations of this phenomenon for a system of nuclear spins
inside different paramagnetic materials [120--127]. The collective relaxation
time of this ultrafast coherent process is inversely proportional to the
number of spins, $N$, and the intensity of magnetodipole radiation is
proportional to $N^2$, in the same way as cooperative radiation time and
radiation intensity of $N$ resonant atoms depend on this number in optic
superradiance [1,29,30,42,45,59]. This is why the process of collective
coherent relaxation of spins has been called {\it superradiant spin
relaxation} or, for short, {\it spin superradiance}. In the case of spin
systems, what is usually measured is not the radiation intensity itself,
which is rather weak, but the power of current induced in the resonant
circuit [128]. The enhancement of generated pulses by using resonators is,
actually, well known in laser optics and is important for realizing
superradiance of Rydberg atoms [129] and recombination superradiance in
electron--hole or electron--positron plasmas [130]. Resonators can be employed
for modifying radiated pulses in optical superradiance [131]. Note also
the usage of resonators for amplifying the nuclear spin echo signals in
magnets [132,133].

The appearance of strong correlations between spins is due to the resonator
feedback field, but not to the photon exchange as it happens for atomic
systems. Hence, various quantum effects existing in the interaction of
electromagnetic field with atoms [32,134--137] seem to be absent in the case
of spin systems. Therefore it looked natural to try, for the theoretical
description of relaxation in a spin system coupled with a resonator, to
invoke the classical Bloch equations complimented by the Kirchhoff equation
for the resonant electric circuit [1,119,138--140]. However, these equations
can provide a description of coherent spin relaxation only when the latter is
triggered by a coherent pulse, similarly to the semiclassical Bloch--Maxwell
equations in optics [1,141,142]. The phenomenon of the self--organized
coherent spin relaxation cannot be described by the Bloch--Kirchhoff
equations. Then, what initiates spin motion leading to the appearance of
purely self--organized spin superradiance? This problem of the origin of
pure spin superradiance was posed by Bloembergen and Pound [119]. They also
noticed that the thermal Nyquist noise of resonator cannot be a mechanism
triggering the motion of spins, since the thermal relaxation time is
proportional to the number of spins in the sample and, thus, the thermal
damping is to be negligibly small for macroscopic samples. Nevertheless, this
notice was forgotten by the following researchers  who assumed that it is just
the thermal noise of resonator which triggers the spin motion.

To resolve this controversy and to discover the genuine mechanisms originating
the spin motion, it was necessary to turn to microscopic models. The system of
nuclear spins is characterized [143] by the Hamiltonian
\be
\label{206}
\hat H = \frac{1}{2} \; \sum_{i\neq j} \; H_{ij} -
\mu_n \; \sum_i\; \vec B\cdot \vec I_i \; ,
\ee
in which spins interact through the dipole potential
$$
H_{ij} = \frac{\mu_n^2}{r_{ij}^3}\;  \left [
\vec I_i\cdot \vec I_j - 3\left (\vec I_i\cdot \vec n_{ij}\right )
\left ( \vec I_j\cdot \vec n_{ij}\right ) \right ] \; ,
$$
where $\mu_n$ is the nuclear magnetic moment, $\vec I_i$ is a nuclear spin
operator, $r_{ij}=|\vec r_{ij}|,\; \vec r_{ij}=\vec r_i - \vec r_j,\; 
\vec n_{ij}=\vec r_{ij}/r_{ij}$. The total magnetic field
$$
\vec B = H_0\; \vec e_z + H\; \vec e_x
$$
contains an external magnetic field $H_0$ and a resonator feedback field
$H$ defined by the Kirchhoff equation.

The temporal behaviour of a finite number of spins, with $27\leq N\leq 343$,
was analysed numerically by computer simulations [144-149]. From various
cases studied, we present here some that give the general qualitative
understanding of the whole picture. In Figs. 1-4, $K_{coh} \equiv
P_{coh}/P_{inc}$ is a coherence coefficient, being the ratio of the
coherent part of the current power $P$ to its incoherent part, and $p_z$ is 
the negative spin polarization. In Figs. 5--11, $C_{coh}\equiv 
I_{coh}/I_{inc}$ is the coherence coefficient of the
average magnetodipole radiation defined as in Eq. (90), with respect to the
total radiation intensity $I$. The current power and radiation intensity are
given in dimensionless units and time is measured in units of $T_2$. In the
figure captions, $p_z(0)$ and $p_x(0)$ mean the corresponding polarization
components at the initial time, $\om_0$ is the Zeeman frequency, $\om$ is
the natural frequency of the resonant electric circuit and also a frequency
of an alternating magnetic field, if any, the amplitude of the latter being
denoted by $h_0$. The quantity
\be
\label{207}
g\equiv \pi^2\eta \; \frac{\rho_n\mu_n^2\om_0}{\hbar\Gm_2\om}
\ee
is the effective coupling parameter, in which $\eta$ is a filling factor;
$\rho_n$, nuclear density; and $\Gm_2=T_2^{-1}$ is a line width. Computer
simulations proved that pure spin superradiance does exist with no thermal
noise involved.

However, computer simulations can provide only a qualitative picture, as the
number of spins considered in such simulations is incomparably smaller than
what one has in real macroscopic samples. Moreover, these simulations give no
analytical formulas, making it difficult, if possible, to classify all
relaxation regimes occurring when varying the numerous parameters of the
system. Simplified models [150] can also provide only a qualitative
understanding.

An analytical solution of the evolution equations corresponding to the
microscopic Hamiltonian (206) and a complete analysis of different
relaxation regimes of nonequilibrium nuclear magnets coupled with a
resonator has been done [25,26,151-158] by employing the scale separation
approach. The evolution equations are written for the averages
\be
\label{208}
u\equiv\frac{1}{N} \; \sum_i < S_i^-> \; , \qquad
s\equiv \frac{1}{N}\; \sum_i< S_i^z>\; ,
\ee
where $S_i^-=S_i^x-iS_i^y$. Presenting local fluctuating fields through
stochastic variables $\xi_0$ and $\xi$, one comes [25,26] to the evolution
equations
$$
\frac{du}{dt} = i\left ( \om_0 -\xi_0 + i\Gm_2\right ) \; u -
i( \gm_3h+\xi) \; s \; ,
$$
\be
\label{209}
\frac{ds}{dt} = \frac{i}{2} \left ( \gm_3 h + \xi\right )\; u^* -
\frac{i}{2} \left ( \gm_3 h + \xi^*\right ) \; u -
\Gm_1 ( s -\zeta)\; ,
\ee
$$
\frac{d}{dt} \; |u|^2 = - 2\Gm_2 |u|^2 - i \left ( \gm_3 h +\xi\right )\;
su^* + i \left (\gm_3 h + \xi^* \right )\; su \; ,
$$
in which the resonator feedback field, $h$, in dimensionless units, satisfies
the Kirchhoff equation
\be
\label{210}
\frac{dh}{dt}  + 2\gm_3 h +\om^2\int_0^t h(t') \; dt '= -
2\kappa\; \frac{d}{dt} \left ( u^* + u\right ) + \gm_3 f \; ,
\ee
in which $f$ is an electromotive force, $\gm_3$ is the resonator ringing
width, and $\kappa\equiv\pi\eta\rho_n\mu_n^2/\hbar\gm_3$. The random local
fields are defined as Gaussian stochastic variables with the stochastic
averages
\be
\label{211}
\ll \xi_0^2\gg \; =\; \ll|\xi|^2\gg\; = \Gm_*^2 \; ,
\ee
where $\Gm_*$ is the inhomogeneous dipole broadening. Because of the existence
of the small parameters
\be
\label{212}
\frac{\Gm_1}{\om_0} \ll 1\; , \qquad \frac{\Gm_2}{\om_0}\ll 1 \; , \qquad
\frac{\Gm_*}{\om_0} \ll 1\; , \qquad \frac{\gm_3}{\om_0}\ll 1 \; , \qquad
\frac{|\Dlt|}{\om_0} \ll 1\; ,
\ee
where $\Dlt\equiv\om-\om_0$, the functions $u$ and $h$ can be classified as
fast while $s$ and $|u|^2$ as slow.

Solving Eqs. (209) and (210), it was shown [25,26,151] that the role of the
thermal Nyquist noise in starting the relaxation process is negligible. But
the main cause triggering the motion of spins, leading to coherent
self--organization, is the action of {\it nonsecular dipole interactions}.
This gives the answer to the question posed by Bloembergen and Pound [119]:
what is the origin of self--organized coherent relaxation in spin systems?
All possible regimes of nonlinear spin dynamics have been analysed. When
the nonresonant external pumping is absent, that is $\zeta>0$, there are
seven qualitatively different transient relaxation regimes: {\it free
induction, collective induction, free relaxation, collective relaxation,
weak superradiance, pure superradiance}, and {\it triggered superradiance}
[25,26,151,155].

In the presence of pumping, realized e.g. by means of dynamical nuclear
polarization directing nuclear spins against an external constant magnetic
field, one has $\zeta\leq 0$. Then, as was shown using phenomenological
equations [139], two stationary solutions can appear. In our approach, the
behaviour of the system is as follows [158]. When $\zeta\leq 0$, three
dynamical regimes can be observed, depending on the value of $\zeta$ with
respect to the pump thresholds
\be
\label{213}
\zeta_1 = -\frac{1}{g} \; , \qquad
\zeta_2 = -\frac{1}{g} \left ( 1 +\frac{\Gm_1}{8\Gm_2}\right ) \; .
\ee
Analysing the equations for the slow variables $s$ and $w$, where
\be
\label{214}
w\equiv |u|^2 - \frac{\Gm_*^2}{\om_0^2} \; s^2 \; ,
\ee
we find two fixed points
\be
\label{215}
s_1^* = \zeta \; , \quad w_1^* = 0\; ; \qquad s_2^* = -\; \frac{1}{g}\; ,
\quad w_2^* = -\frac{\Gm_1(1+g\zeta)}{\Gm_2 g^2} \; .
\ee
When $\zeta_1<\zeta\leq 0$, the first fixed point is a stable node and the
second one is a saddle point. For $\zeta=\zeta_1$, both points merge together,
being neutrally stable. After the bifurcation at $\zeta=\zeta_1$, in the
region $\zeta_2\leq\zeta<\zeta_1$, the first fixed point looses its stability
becoming a saddle point while the second fixed point becomes a stable node. 
Finally, when $\zeta<\zeta_2$, the second fixed point  turns into a stable
focus, and the first one continues to be a saddle point. In this way, there
are three qualitatively different lasting relaxation regimes induced by the
pumping [158]. The first one is the {\it incoherent monotonic relaxation}
to the first stationary solution $s_1^*,\; w_1^*$. The second regime is the
{\it coherent monotonic relaxation} to the second stationary solution
$s_2^*,\; w_2^*$, although the level of coherence may be rather low. And the
third case is the {\it coherent pulsing relaxation} to the second fixed
point. This unusual regime of pulsing relaxation was observed experimentally
[159]. Here we present the results of numerical solution of the evolution
equations for the slow variables $s=z(t)$ and $w(t)$ defined in Eq. (214).
Different cases of the pulsing regime are clearly demonstrated in Figs. 12
to 18. In the corresponding figure captions we use the notation
$z_0=z(0),\; w_0=w(0)$, and $\gm\equiv\gm_1/\gm_2$. Everywhere in Figs. 12
to 17, the pump parameter is $\zeta=-0.5$, and in Fig. 18 this parameter is
varied. The coupling parameter (207) is always $g=10$.

The problem of superradiant spin relaxation can be generalized to the case
of nuclei incorporated into a ferromagnetic matrix, where nuclear and
electron spins interact through hyperline forces. Some model studies of this
case have been undertaken [160--162], and a general microscopic theory has
also been developed [163]. The latter theory makes it possible to discover
all feasible causes triggering the process of self--organized coherent
relaxation. The most important such causes are the {\it dipole hyperfine
interactions, dipole nuclear interactions}, and the {\it transverse
magnetocrystalline anisotropy}.

\section{Negative Electric Current}

The study of electric processes in semiconductors is important for describing
and modelling semiconductor devices [164]. One of the most difficult problems
is the consideration of strongly nonequilibrium phenomena in essentially
nonuniform semiconductors. Nonequilibrium and nonuniform distributions of
charge carriers can be formed in several ways, for instance, by means of
external irradiation [165,166]. Transport properties of semiconductors with
essentially nonuniform distribution of charge carriers can be rather specific.
For example, in a sample, biased with an external constant voltage, the
resulting electric current may turn against the latter displaying the
transient effect of negative electric current [3,166--168].

Transport properties of semiconductors are usually described by the
semiclassical drift--diffusion equations [164]. In what follows a plane
device, of area $A$ and length $L$ is considered, which is biased with a
constant voltage $V_0$. It is convenient to pass to dimensionless
quantities, measuring the space variable $x$ in units of $L$, time in
units of the transit time
$$
\tau_0 \equiv \frac{L^2}{\mu V_0} \; , \qquad
\mu\equiv \min\{ |\mu_i|\} \; ,
$$
where $\mu_i$ is a mobility of the $i$--type carriers. And the characteristic
quantities
$$
\rho_0 \equiv\frac{Q_0}{AL} \; , \qquad Q_0\equiv \ep AE_0\; , \qquad
E_0 \equiv \frac{V_0}{L} \; ,
$$
$$
j_0 \equiv\frac{Q_0}{A\tau_0} \; , \qquad D_0\equiv \mu V_0\; , \qquad
\xi_0 \equiv \frac{\rho_0}{\tau_0} \; ,
$$
are employed for measuring other physical values which are used below.

The drift--diffusion equations consist of the continuity equations
\be
\label{216}
\frac{\prt\rho_i}{\prt t} +\mu_i\; \frac{\prt}{\prt x} 
\left ( \rho_i E\right ) - D_i \; \frac{\prt^2\rho_i}{\prt x^2} +
\frac{\rho_i}{\tau_i} = \xi_i\; ,
\ee
for each type of charge carriers, and of the Poisson equation
\be
\label{217}
\frac{\prt E}{\prt x} = 4\pi\sum_i\rho_i
\ee
for the electric field $E(x,t)$. Here $\rho_i(x,t)$ is a charge density;
$\mu_i,\; D_i$, and $\tau_i$ are mobility, diffusion coefficient, and
relaxation time, respectively; $\xi_i$ is a generation--recombination noise
[169]. The sample is biased with an external constant voltage, which in
our dimensionless notation implies that
\be
\label{218}
\int_0^1 E(x,t)\; dx = 1\; .
\ee
At the initial time, the distribution of charge carriers
\be
\label{219}
\rho_i(x,0) = f_i(x)
\ee
is assumed to be nonuniform.

The total electric current through the semiconductor sample is
\be
\label{220}
J(t) \equiv \int_0^1 j(x,t)\; dx\; ,
\ee
where the density of current
\be
\label{221}
j = \sum_i\left ( \mu_i E - D_i\; \frac{\prt}{\prt x}\right ) \rho_i +
\frac{1}{4\pi} \; \frac{\prt E}{\prt t} \; .
\ee

Because of the voltage integral (218), one has
\be
\label{222}
\int_0^1 \frac{\prt}{\prt t}\;  E(x,t) \; dx = 0 \; .
\ee
It is also possible to show that
\be
\label{223}
\lim_{\tau\ra\infty} \ll \frac{1}{\tau}\; \int_0^\tau 
\frac{\prt}{\prt x}\; E(x,t) \; dt \gg \; = 0 \; .
\ee
This means that the function $E$ can be considered as slow on average in
time and in space. Then, treating $E$ as a quasi--invariant, one may find the
solutions to Eqs. (216) and (217) in order to analyse their general
space--time behaviour and to find conditions when the effect of negative
electric current could arise. Such negative current can appear only when
the initial charge distribution is essentially nonuniform. For example, if
this initial charge distribution forms a narrow layer located at the point
$x=a$, then the total current (220) becomes negative for a transient interval
of time in the vicinity of $t=0$, if one of the following conditions holds
true:
\be
\label{224}
a< \frac{1}{2} -\frac{1}{4\pi Q} \quad
\left ( Q > \frac{1}{2\pi} \right )\; , \quad {\rm or} \quad
a > \frac{1}{2} +\frac{1}{4\pi |Q|} \quad
\left ( Q < -\frac{1}{2\pi} \right )\; ,
\ee
where
$$
Q \equiv \sum_i Q_i \; , \qquad Q_i \equiv\int_0^1 \rho_i(x,0) \; dx \; .
$$
The effect of the negative electric current can be employed for
various purposes, as is discussed in Refs. [3,168]. For instance, when the
initial charge layer is formed by an ion beam irradiating the semiconductor
sample, the location $a$ corresponds to the ion mean free path. In this case,
by measuring the negative current $J(0)$, one can define this mean free path
\be
\label{225}
a = \frac{1}{2} - \frac{1}{4\pi Q} \left [ 1  -
\frac{J(0)}{\sum_i\mu_i Q_i}\right ] \; .
\ee
This formula is valid for both positive and negative values of $Q$.

Equations (216) and (217) have also been solved numerically [3,168], which
confirmed the appearance of the negative electric current. Two cases were
analysed, with one layer of charge carriers and with two such layers. Here
we present the results of calculations for the double--layer case. The
initial charge distributions (219) are given by the Gaussians
$$
f_i(x) =\frac{Q_i}{Z_i} \exp\left\{-\;\frac{(x-a_i)^2}{2b_i}\right\} \; ,
$$
in which $0\leq a_i\leq 1$ and
$$
Q_i = \int_0^1 f_i(x)\; dx \; , \qquad
Z_i =\int_0^1 \exp\left\{ -\; \frac{(x-a_i)^2}{2b_i}\right\} \; dx \; .
$$
The positive charge carriers, with $\mu_1=1$ and $Q_1=1$, form the left layer
centered at $a_1=a$, while the negative charge carriers form the layer
centered at $a_2=1-a$. We keep in mind the relation $D_2=3D_1$ for the
diffusion coefficients, typical for holes and electrons, and we set
$D_1=10^{-3}$. For short, we use the notation $\tau_1^{-1}=\tau_2^{-1}=\gm$
and $b_1=b_2=b$. The generation--recombination noise is neglected, which is
admissible at the initial stage of the process. As the boundary conditions,
we accept the absence of diffusion through the semiconductor surface, which
implies the Neumann boundary condition
$$
\frac{\prt}{\prt x}\; \rho_i(x,t) = 0 \qquad (x=0,\; x=1)\; .
$$
In Figs. 19 to 24, we present the total current (220) as well as the electric
current through the left surface, $J(0,t)\equiv j(0,t)$ and through the right
surface, $J(1,t) \equiv j(1,t)$, defined by the current (221) at $x=0$ or
$x=1$, respectively.

\section{Magnetic Semiconfinement of Atoms}

Dynamics of neutral atoms in nonuniform magnetic fields concerns problems of
current experimental and theoretical interest. By means of such fields, atoms
can be confined inside magnetic traps, which allows to accomplish various
experiments with the systems of trapped atoms. Recently, Bose--Einstein
condensation has been attained in a dilute gas of trapped atoms of $^{87}$Rb
[170], $^7$Li [171], Na [172], and H$\downarrow$ [173]. The details on
theory and experiment can be found in reviews [174--176]. The Bose--Einstein
condensate is believed to form, at least partially, a coherent state. If it
would be possible to construct a device emitting a coherent atomic beam, this
would be analogous to a laser radiating a coherent photon ray. This is why
one may call the device, emitting a coherent atomic beam, an atom laser
[177--184]. An output coupler, coherently extracting condensed atoms form
a trap, was demonstrated recently [185--187]. But in these demonstrations,
the atoms, when escaping from a trap, fly out more or less in all directions,
with anisotropy formed only by the gravitational force. While the very
first condition on a laser is that its output is highly directional, with
the possibility of varying the beam direction [183].

A mechanism for creating well--collimated beams of neutral atoms was
advanced in Refs. [188--192]. This mechanism suggests an output coupler
that extracts trapped atoms in the form of a directed beam.

The motion of neutral atoms in magnetic fields can be described by the
semiclassical equations for the quantum--mechanical average of the real--space
coordinate $\vec r=\{ r_\al\}$, where $\al=x,y,z$, and for the average
$\vec S=\{ S_\al\}$ of the spin operator [193--195]. The first equation writes
\be
\label{226}
m\; \frac{d^2r_\al}{dt^2} = \mu_0\; \vec S\cdot
\frac{\prt\vec B}{\prt r_\al} + mg_\al + f_\al \; ,
\ee
where $m$ and $\mu_0$ are mass and magnetic moment of an atom; $\vec B$ is a
magnetic field; $g_\al$ is a component of the standard gravitational
acceleration; and $f_\al$ is a collision force component. The equation for
the average spin is
\be
\label{227}
\hbar \; \frac{d\vec S}{dt} = \mu_0\;  \vec S\times\vec B \; .
\ee
The total magnetic field
$$
\vec B =\vec B_1 + \vec B_2\; ,
$$
\be
\label{228}
\vec B_1 = B_1'\left ( x\;\vec e_x + y\;\vec e_y + 
\lbd z\;\vec e_z\right ) \; , \qquad
\vec B_2 = B_2 \left ( h_x \;\vec e_x + h_y\; \vec e_y\right ) \; ,
\ee
where $|\vec h|=1$, consists of the quadrupole field $\vec B_1$, typical of
quadrupole magnetic traps, and of a transverse field, e.g., of a rotating
field [196,197]. In the quadrupole field, $\lbd$ is the anisotropy parameter.

It is convenient to pass to the dimensionless space variable, measuring the
components of $\vec r$ in units of the characteristic length
\be
\label{229}
R_0 \equiv\frac{B_2}{B_1'} \; .
\ee
Introduce the characteristic frequencies by the relations
\be
\label{230}
\om_1^2 \equiv \; \frac{\mu_0B_1'}{mR_0} \; , \qquad
\om_2 \equiv \; \frac{\mu_0B_2}{\hbar}\; , \qquad
\om\equiv \max_t\left |\frac{d\vec h}{dt}\right | \; .
\ee
Also, we define
\be
\label{231}
\dlt_\al \equiv \frac{g_\al}{R_0\om_1^2}\; , \qquad
\gm\xi_\al \equiv\frac{f_\al}{mR_0} \; ,
\ee
where $\gm$ is a collision rate and $\xi_\al$ can be treated as a random
variable with the stochastic averages
$$
\ll\xi_\al(t)\gg\; = 0\; , \qquad
\ll\xi_\al(t)\;\xi_\bt(t')\gg \; =2D_\al\; \dlt_{\al\bt}\; \dlt(t-t') \; ,
$$
in which $D_\al$ is a diffusion rate. Then Eq. (226) can be written as the
stochastic differential equation
\be
\label{232}
\frac{d^2\vec r}{dt^2} = \om_1^2 \left ( S_x\;\vec e_x + S_y\;\vec e_y +
\lbd S_z\; \vec e_z +\vec\dlt\right ) +\gm\vec\xi \; ,
\ee
and Eq. (227) acquires the form
\be
\label{233}
\frac{d\vec S}{dt} = \om_2\; \hat A\; \vec S\; ,
\ee
in which the antisymmetric matrix $\hat A=[A_{\al\bt}]$ has the elements
$$
A_{\al\bt} = -A_{\bt\al}\; , \qquad A_{\al\al} = 0 \; ,
$$
$$
A_{12} = \lbd z\; , \qquad A_{23} = x + h_x \; , \qquad A_{31}= y+h_y\; .
$$

Assuming the occurrence of the small parameters
\be
\label{234}
\left |\frac{\gm}{\om_1}\right | \ll 1 \; , \qquad
\left |\frac{\om_1}{\om_2}\right | \ll 1 \; , \qquad
\left |\frac{\om}{\om_2}\right | \ll 1 \; ,
\ee
we may classify the variables $\vec r$ and $\vec h$ as slow, compared to the
fast spin variable $\vec S$. Then Eq. (233) can be solved yielding
\be
\label{235}
\vec S(t) =\sum_{i=1}^3\; a_i\;\vec b_i(t)\; \exp\{ \bt_i(t)\} \; ,
\ee
where
$$
a_i= \vec S(0)\cdot\vec b_i(0) \; ,
$$
$$
\vec b_i(t) =\frac{1}{\sqrt{C_i}}\left [\left ( A_{12}A_{23} -\al_i A_{31}
\right )\vec e_x +\left ( A_{12}A_{31} +\al_i A_{23}\right )\vec e_y +
\left ( A_{12}^2 +\al_i^2\right )\vec e_z\right ] \; ,
$$
$$
C_i =\left ( A_{12}^2 - |\al_i|^2\right )^2 +
\left ( A_{12}^2 +|\al_i|^2\right )
\left ( A_{23}^2 + A_{31}^2\right ) \; ,
$$
$$
\al_{1,2} =\pm i\al \; , \qquad \al_3 = 0 \; , \qquad
\al^2 \equiv A_{12}^2 + A_{23}^2 + A_{31}^2 \; , \qquad
\bt_i(t) = \om_2\int_0^t \al_i(t')\; dt'\; .
$$
Substituting  Eq. (235) into the right--hand side of Eq. (232) and
averaging the latter over time and over stochastic variables, we obtain
\be
\label{236}
\frac{d^2\vec r}{dt^2} =\vec F +\om_1^2\; \vec\dlt \; ,
\ee
where
$$
\vec F\equiv \om_1^2 a_3 < b_3^x\;\vec e_x + b_3^y\;\vec e_y +
\lbd b_3^z\; \vec e_z >\; ,
$$
$$
a_3 = \frac{(x+h_x^0)S_x^0+(y+h_y^0)S_y^0+\lbd zS_z^0}
{[(x+h_x^0)^2+(y+h_y^0)^2+\lbd^2 z^2]^{1/2}} \; , \qquad
\vec b_3 = \frac{(x+h_x)\;\vec e_x+(y+h_y)\;\vec e_y+\lbd z\;\vec e_z}
{[(x+h_x)^2+(y+h_y)^2+\lbd^2 z^2]^{1/2}} \; ,
$$
angle brackets imply time averaging and
$h_\al^0\equiv h_\al(0),\; S_\al^0\equiv S_\al(0)$. For the rotating
transverse field, with
\be
\label{237}
h_x =\cos\om t \; , \qquad h_y =\sin\om t\; ,
\ee
we find
$$
\vec F = \frac{\om_1^2[(1+x)S_x^0 + yS_y^0 +\lbd z S_z^0]\;
(x\;\vec e_x + y\;\vec e_y +2\lbd^2 z\;\vec e_z)}
{2[(1+2x+x^2 + y^2+\lbd^2z^2)(1+x^2+y^2+\lbd^2z^2)]^{1/2}} \; .
$$

The motion of atoms, described by Eq. (236), essentially depends on the
initial state, which, as is known [198,199], can be prepared in an arbitrary
way. Suppose that atoms, after being laser cooled in a magneto--optical
trap [200], are loaded into a magnetic trap where they are further cooled by
evaporative cooling down to sufficiently low temperatures, so that there is
a portion of atoms with low velocities, which are located close to the trap
center. If the initial spin condition for these atoms is such that $S_x^0<0$
and $S_y=S_z=0$, then the atoms are confined inside the trap moving in an
approximately harmonic potential. The gradient of the quadrupole field
supplies the levitating force to support atoms against gravity. The
combination of the magnetic field and gravity produces a very nearly
harmonic confining potential within the trap
volume in all three dimensions [201].

The semiconfining regime of motion [188--192] can be realized by preparing
for the spin variable nonadiabatic initial conditions
\be
\label{238}
S_x^0=S_y^0 = 0 \; , \qquad S_z^0 \equiv S\neq 0\; .
\ee
Such conditions can be arranged in several ways. One possibility could be to
confine atoms in a trap, where all atoms are polarized having their spins
in the $z$ direction, as e.g. in the trap of Ref. [201], being a quadrupole
trap with a bias field along the $z$ axis. Then the longitudinal bias field is
quickly switched off, and at the same time, a transverse field is switched on,
which would correspond to the sudden change of potential [202]. Another way
could be to prepare spin polarized atoms in one trap and quickly load them
into another trap with the required field configuration. Atoms can be
prepared practically $100\%$ polarized [203], with the spin--spin
relaxation time reaching $100$ s [204]. The possibility of realizing two
ways of transferring atoms from one trap to another, by means of sudden
transfer as opposed to adiabatic transfer, is discussed in Ref. [205]. The
third way of organizing the nonadiabatic initial conditions (238) could be
by acting on the trapped atoms with a short pulse of strong magnetic
field, polarizing atomic spins in the desired way.

With the initial conditions (238), the motion of atoms becomes axially
restricted from one side, depending on the sign of $\lbd S$. Atoms fly out
of the trap predominantly in one direction, forming a well--collimated beam
[188--192]. This mechanism can be used for atom lasers. Another possibility
could be to study the dynamics of binary mixtures of Bose systems, where
the effect of conical stratification [206] can arise. The mixtures of two
condensates have been realized for rubidium [207] and sodium [208], and the
dynamics of two rubidium condensates was observed in Ref. [209].

When solving equation (236) for the realistic case of a finite trap, one
should take into account the trap shape factor, which can be written in the
Gaussian form
$$
\vp(\vec r) =\exp\left ( -\; \frac{x^2+y^2}{R^2} -
\frac{z^2}{L^2} \right ) \; ,
$$
where $R$ and $L$ are the trap radius and length. The relation between the
latter can be quite different for different traps, starting from almost
spherical traps, where $R\approx L$, to needle--shape traps, with $R/L\sim
10^{-3}$, as for Ioffe--Pritchard magnetic traps [210]. Accepting the initial
spin conditions (238), and using the notation
$$
f(\vec r) \equiv \frac{\vp(\vec r)}{[(1+2x+x^2+y^2+\lbd^2z^2)
(1+x^2+y^2+\lbd^2z^2)]^{1/2}} \; ,
$$
from Eq. (236) we obtain
\be
\label{239}
\frac{d^2x}{dt^2} = \om_1^2 \left (\frac{\lbd}{2}\; S\; f\; z\; x +\dlt_x
\right ) \; , \qquad
\frac{d^2z}{dt^2} = \om_1^2\left ( \lbd^3 S\; f\; z^2 +\dlt_z\right ) \; ,
\ee
where the equation for $y$, being similar to that for $x$, is not written
down. Note that instead of the Gaussian shape factor for the trap, one could
opt for
$$
\vp(\vec r) = 1 -
\Theta(x-R)\;\Theta(y-R)\; \Theta\left ( |z| -\frac{1}{2}\; L\right )\; ,
$$ 
with $\Theta(\cdot)$ being the unit--step function.

Equations (239) were analysed both analytically and numerically [188--192].
Their solutions display the semiconfined regime of motion. Taking into
account random pair collisions in Eq. (232) shows that atomic collisions do
not disturb the semiconfined motion provided that temperature $T$ is
sufficiently low, satisfying the condition
\be
\label{240}
\frac{k_BT\hbar\rho^2 a_s^2}{m^2\om_1^3}\; \ll 1 \; ,
\ee
in which $\rho$ is the density of atoms and $a_s$, their scattering length.
The semiconfined regime of motion makes it possible to form well--collimated
beams on neutral atoms by means of only magnetic fields.

\section{Nuclear Matter Lasing}

The natural question that arises after talking about atom lasers is whether
there can be produced matter waves corresponding to other Bose particles,
which could be employed for lasing. One such possibility is related to the
creation of large number of pions in hadronic, nuclear, and heavy--ion
collisions. If the density of pions appearing in the course of these
collisions is sufficiently high, then correlations between pions can result
in the formation of coherent state and in the feasibility of realizing a
pion laser [211]. Pions are not the sole type of Bose particles arising in
nuclear matter under extreme conditions characteristic of fireballs produced
in high--energy collisions [212,213]. There are plenty of reviews devoted to
the state of nuclear matter at extreme conditions, including the region
of deconfinement transition. Here we cite only some recent of such reviews
[214--217].

The very first necessary condition that is required for lasing is to be able
to generate Bose particles with sufficiently high density. Therefore, in
order to answer the question what kind of Bose particles appearing in nuclear
matter under extreme conditions could be used for lasing, one has, first of
all, to find out what are these Bose particles and under what conditions
their density is maximal. In this section, we give a very brief account of
an analysis based on the multichannel model of nuclear matter [217--221].
The main idea in constructing this model goes back to the Weinberg approach
for describing composite particles [222--224], with effective Hamiltonians
that are assumed to be a result of the Fock--Tani transformation [225]. Now
we shall not plunge into the details of the multichannel model, which can be
found in Refs. [217,219], but we shall present some figures and will formulate
the conclusion of an analysis [221] with regard to the most probable
candidates for nuclear matter lasing.

When rising temperature or density, nuclear matter exhibits a transition
from hadron state to quark--gluon state. This transition is often assumed
to be a sharp first--order transition. Lattice numerical simulations for the
quarkless $SU(3)$ gauge model show that deconfinement is really a first--order
phase transition [226], which is in agreement with the multichannel model.
Figures 25 to 27 illustrate the behaviour of some thermodynamic
characteristics, normalized to the corresponding Stefan--Boltzmann limits,
for the case of the $SU(3)$ gluon--glueball mixture. Figure 28 shows the
related glueball channel probability. The sharpness on the deconfinement
transition essentially depends on the interactions between particles or on
their radii, when the composite particles are treated as bags [227].

In the case of realistic nuclear matter, deconfinement is rather a gradual
crossover but not a genuine phase transition [217]. Then all thermodynamic
characteristics change continuously, without jumps. This concerns as well the
channel probabilities. Thus, in Figs. 29, 30 the channel probabilities of
nucleons and dibaryons are shown as functions of baryon density normalized
to the normal baryon density of nuclear matter $n_{0B}=0.167$ fm$^{-3}$. The
possible appearance of dibaryons is of special interest since they, being
bosons, can form a Bose condensate [217,228--230].

Summarizing the results of the analysis [221], three types of Bose particles
can appear in nuclear matter in large quantities: pions, dibaryons, and
gluons. The maximum of the pion channel probability, reaching $w_\pi =0.6$,
occurs in the vicinity of the deconfinement transition at $T\approx 160$ MeV
and low baryon densities $n_B<n_{0B}$. Dibaryons can appear mainly at low
temperatures $T<20$ MeV and relatively high baryon densities
$n_B\sim 10\; n_{0B}$, where their channel probability $w_6\approx 0.7$.
Large amount of gluons emerges only at high temperatures $T>160$ MeV.  
In addition, one should keep in mind that gluons cannot be observed as
free particles.

Talking about possible pion, dibaryon, or gluon lasing from nuclear matter,
we have touched here just one necessary condition, trying to find out when
these Bose particles can appear in large quantities. To realize such a lasing
in reality will, certainly, require to solve a number of other problems.
But, anyway, to understand the conditions when this lasing could be plausible
in principle is the necessary first step.

\section{Conclusion}

We have described a general method for treating strongly nonequilibrium
processes in statistical systems. This method is called the {\it Scale
Separation Approach} since its basic idea is to try to separate different
characteristic scales of time and space variables. The idea itself is,
of course, not new and we have employed some known techniques. What is
original in our approach is: (i) The combination of several methods and
their adjustment to the problems of nonequilibrium statistical mechanics.
(ii) The generalization of the averaging method to stochastic and partial
differential equations. (iii) Probabilistic solution of the problem of
pattern selection.

The scale separation approach has been shown to be very useful for describing
cooperative phenomena in the interaction of radiation with matter. To
emphasize the generality of the approach, it is illustrated here by several
different physical examples, whose common feature is that the related
evolution equations are nonlinear differential or integro--differential
stochastic equations. Such equations, as is known, are difficult to solve.
The scale separation approach makes it possible to find accurate approximate
solutions. The accuracy of these solutions has been confirmed by numerical
calculations and by comparison with experiment, when available. Using this
approach, several interesting physical problems have been solved and new
effects are predicted. Among the most interesting applications we would 
like to emphasize the following.

{\it Collective Liberation of Light} happens when en ensemble of resonant 
atoms is doped into a medium with polariton band gap. If the transition 
frequency of an atom is inside this prohibited gap, then atomic 
spontaneous emission is strongly suppressed, which is termed localization 
of light. Although spontaneous emission of a single atom is prohibited, a 
collective of such atoms can radiate due to their coherent interactions. 
As a result of this coherent radiation, light becomes partially 
liberated. We have advanced dynamical theory of this light liberation for 
the realistic situation when the radiation wavelength is smaller than the 
linear sizes of the sample (see Sec. 6).

{\it M\"ossbauer Magnetic Anomaly} has puzzled researches for many years. 
This anomaly consists in a strong increase of the area under the 
M\"ossbauer spectrum, below the temperature of magnetic phase transition. 
Several explanations of this anomaly have been suggested. We have thoroughly
analysed this phenomenon and concluded that previously suggested 
mechanisms cannot explain this anomaly but that its origin is rather in 
the increase of inhomogeneous broadening of M\"ossbauer nuclei, which is 
due to the arising magnetic field (see Sec. 8).

{\it Turbulent Photon Filamentation} in resonant media is an intriguing
example of self-organization in a strongly nonequilibrium system, whose 
dynamical theory was absent. We have developed such a theory, based on 
the probabilistic approach to pattern selection, and showed that it gives 
agreement with experiment (see Sec. 10).

{\it Superradiant Spin Relaxation} occurs in a system of spins coherently 
interacting with each other through resonator feedback field. This 
ultrafast coherent relaxation is similar to superradiance in optical 
systems, because of which the term spin superradiance was coined. 
Contrary to its optical counterpart, the origin of purely self-organized 
spin superradiance has not been understood for about 40 years, after 
Bloembergen and Pound posed this problem in 1954. We have developed a 
theory of nonlinear spin dynamics, based on a microscopic Hamiltonian, 
elucidated the origin of pure spin superradiance, and described all main 
regimes of spin relaxation, without pumping as well as in the presence of 
the latter (see Sec. 11).

{\it Negative Electric Current} is a rather unusual effect, when electric 
current flows against an applied voltage. This is a transient effect that 
can occur in nonuniform semiconductors. We have predicted this effect and 
suggested its theory (see Sec. 12).

{\it Magnetic Semiconfinement of Atoms} is another effect we predict. 
This effect can serve as a mechanism for creating well--collimated beams 
of neutral atoms by means of magnetic fields. It can be used to form 
coherent beams of Bose atoms from atom lasers. We have presented a theory 
of this effect (see Sec. 13).

The possibility of treating nonequilibrium processes in nonlinear systems 
of quite different nature has become possible owing to the Scale 
Separation Approach, which provides accurate approximate solutions to 
complicated systems of differential and integro--differential equations.

\vskip 5mm

{\bf Acknowledgement}

\vskip 2mm

We are grateful for discussions and useful advice to V.S. Bagnato,
N.A. Bazhanov, C.M. Bowden, M.G. Cottam, V.I. Emelyanov, R. Friedberg,
S.R. Hartmann, V.K. Henner, Vl.V. Kocharovsky, J.T. Manassah, A.N. Oraevsky, 
T. Ruskov, V.V. Samartsev, M.A. Singh, and R. Tanas. We appreciate the 
contribution of all our coauthors.

\newpage

\begin{center}

{\bf Figure Captions}

\end{center}

{\bf Fig.1.}  Coherence coefficient $K_{coh}$, current power $P$, and spin
polarization $p_z$ as functions of time for two different coupling
parameters defined in Eq. (207), $g_1$ (solid line) and $g_2$ (dashed line),
with the relation $g_1/g_2=10$.

\vspace{5mm}

{\bf Fig.2.} The same as in Fig. 1 for two different Zeeman frequencies,
$\om_{01}$ (solid line) and $\om_{02}$ (dashed line), related by the ratio
$\om_{01}/\om_{02}=5$.

\vspace{5mm}

{\bf Fig.3.} The same functions as in Fig. 1 for different initial
polarizations, $p_{z1}(0)$ (solid line) and $p_{z2}(0)$ (dashed line),
with the relation $p_{z1}/p_{z2}(0)=2$.

\vspace{5mm}

{\bf Fig.4.} The same functions as in Fig. 1 for different initial
transverse polarizations, $p_{x1}(0)$ (solid line) and $p_{x2}(0)$ (dashed
line), with the relation $p_{x1}/p_{x2}(0)=0.5$.

\vspace{5mm}

{\bf Fig.5.}  Coherence coefficient $C_{coh}$, radiation intensivity $I$,
and spin polarization $p_z$ versus time for  $p_z(0)=0.48$ and different 
parameters: $\om_0=200,\; g=25$ (solid line); $\om_0=40,\; g=25$ (dashed
line); and $\om_0=40,\; g=2.5$ (solid line with crosses).

\vspace{5mm}

{\bf Fig.6.}  Coherence coefficient $C_{coh}$, radiation intensivity $I$,
and spin polarization $p_z$ as functions of time in the case of
switched--off resonator--spin coupling ($g=0$). The varied parameters
are: $\om_0=200,\; p_x(0)=0.48$ (solid line); $\om_0=20,\; p_x(0)=0.48$
(dashed line); and $\om_0=200,\; p_x(0)=0.20$ (solid line with crosses).

\vspace{5mm}

{\bf Fig.7.} The same as in Fig. 6 for $p_x(0)=0.48$ and for different
Zeeman frequencies: $\om_0=1000$ (solid line); $\om_0=200$ (dashed line);
$\om_0=50$ (solid line with crosses); and $\om_0=200$ with switched--off
dipole interaction (solid line with triangles).

\vspace{5mm}

{\bf Fig.8.} The same as in Fig. 6 for $p_x(0)=0.48$ but in the presence 
of an alternating magnetic field with the frequency $\om=\om_0$ and
different amplitudes: $h_{01}$ (solid line); $h_{02}$ (dashed line); where
$h_{01}/h_{02}=10$; and $h_{03}=0$ (solid line with crosses).

\vspace{5mm}

{\bf Fig.9.} The same as in Fig. 8 but for $p_x(0)=-0.48$ and
different amplitudes of the alternating field: $h_{01}$ (solid line);
$h_{02}$ (dashed line); and $h_{03}$ (solid line with crosses), where the
amplitude relations are $h_{01}/h_{02}=0.25$ and $h_{01}/h_{03}=0.1$. 

\vspace{5mm}

{\bf Fig.10.} The same as in Fig. 8  for  a varying relative detuning from
the resonance $\dlt\equiv(\om-\om_0)/\om_0$ taking the values: $\dlt=0$
(solid line); $\dlt=0.025$ (dashed line); and $\dlt=0.25$ (solid line with
squares).

\vspace{5mm}

{\bf Fig.11.} Radiation intensivity $I$, coherence coefficient $C_{coh}$,
and spin polarization $p_z$ versus time, in the absence of alternating
external fields and with a weak coupling with a resonator, $g\sim 1$.

\vspace{5mm}

{\bf Fig.12.} Phase portrait demonstrating a stable focus for the
parameters $z_0=-0.5$, $w_0=0.001,\; g=10$, and $\gm=0.1$.

\vspace{5mm}

{\bf Fig.13.} Pulsing regime of spin relaxation with the parameters
$z_0=-0.1,\; w_0=10^{-6}$ and $\gm=0.01$ for the functions: (a) $w(t)$;
(b) $z(t)$.

\vspace{5mm}

{\bf Fig.14.} The time dependence of the functions: (a) $w(t)$; (b)
$z(t)$, for the parameters $z_0=-0.5,\; w_0=0.001$, and $\gm=1$.

\vspace{5mm}

{\bf Fig.15.} Dynamics of slow solutions: (a) $w(t)$; (b) $z(t)$, for the
parameters $z_0=-0.5,\; w_0=0.01$, and $\gm=0.1$.

\vspace{5mm}

{\bf Fig.16.} Evolution of slow solutions: (a) $w(t)$; (b) $z(t)$, for the
parameters $z_0=0.5,\; w_0=0.01$, and $\gm=0.01$.

\vspace{5mm}

{\bf Fig.17.} Temporal behaviour of the function $w(t)$ for different sets
of parameters: (a) $z_0=-0.1,\; w_0=10^{-6},\; \gm=0.001$; (b)
$z_0=-0.1,\;\om_0=0.001,\; \gm=0.01$; (c) $z_0=-0.5,\;
w_0=10^{-6},\;\gm=0.1$; (d) $z_0=-0.5,\; w_0=0.001,\gm=0.01$.

\vspace{5mm}

{\bf Fig.18.} Function $w(t)$ for $z_0=0.5,\; w_0=0.5,\; \gm=1$, and
varying pump parameters: $\zeta=-0.5$ (solid line); $\zeta=-0.3$ (dashed
line).

\vspace{5mm}

{\bf Fig.19.} Electric current through the semiconductor surfaces in the
case of $a=0.1,\; Q_2=-1,\;\gm=1$ and different mobilities: $\mu_2=-10$
(solid line); $\mu_2=-5$ (dashed line); $\mu_2=-3$ (short--dashed line).
(a) Left--surface current $J(0,t)$; (b) Right--surface current $J(1,t)$.

\vspace{5mm}

{\bf Fig.20.} Left--surface current $J(0,t)$ (solid line),
right--surface current $J(1,t)$ (dashed line), and the total current
$J(t)$ (short--dashed line) for $a=0.25,\; Q_2=-0.1,\;\mu_2=-10$ and
different relaxation parameters: (a) $\gm=1$; (b) $\gm=10$; (c) $\gm=25$.

\vspace{5mm}

{\bf Fig.21.} Total electric current $J(t)$ for 
$a=0.25,\;Q_2=-0.1,\;\mu_2=-10$ and varying relaxation parameters: $\gm=1$
(solid line); $\gm=10$ (dashed line); $\gm=25$ (short--dashed line).

\vspace{5mm}

{\bf Fig.22.} Electric current through semiconductor for the parameters
$a=0.1,\;Q_2=-1,\;\gm=1$ and different mobilities: $\mu_2=-10$
(solid line); $\mu_2=-5$ (dashed line); $\mu_2=-3$ (short--dashed line).

\vspace{5mm}

{\bf Fig.23.} Electric current $J(t)$ as a function of time for 
$a=0.1,\;\mu_2=-3,\;\gm=1$ and different initial charges:: $Q_2=0$
(solid line); $Q_2=-0.25$ (dashed line); $Q_2=-0.5$ (short--dashed line);
$Q_2=-0.75$ (dotted line); $Q_2=-1$ (dashed--dotted line).

\vspace{5mm}

{\bf Fig.24.} Electric current $J(t)$ for $Q_2=-1,\;\mu_2=-3,\;\gm=1$,
and different locations of initial charge layers: $a=0.05$
(solid line); $a=0.1$ (dashed line); $a=0.15$ (short--dashed line);
$a=0.2$ (dotted line); $a=0.25$ (dashed--dotted line).

\vspace{5mm}

{\bf Fig.25.} Relative energy density as a function of temperature in MeV
for the $SU(3)$ gluon--glueball mixture of different glueball radii: 0
(line 1); 0.5 fm (line 2); 0.7 fm (line 3); 0.8 fm (line 4); 1 fm (line 5).

\vspace{5mm}

{\bf Fig.26.} Relative enthalpy for the gluon--glueball mixture as a
function of temperature reduced to the deconfinement temperature, in the
case of the glueball radius 0.82 fm, compared with the lattice numerical
calculations.

\vspace{5mm}

{\bf Fig.27.} Relative specific heat for the gluon--glueball mixture,
for the glueball radius 0.82 fm, as a function of temperature in MeV.

\vspace{5mm}

{\bf Fig.28.} Glueball channel probability versus temperature in MeV for
the glueball radii as in Fig. 25.

\vspace{5mm}

{\bf Fig.29.} Nucleon channel probability as a function of relative baryon
density.

\vspace{5mm}

{\bf Fig.30.} Dibaryon channel probability versus relative baryon density.

\end{document}